\documentclass[aps,pra,twocolumn,showpacs,superscriptaddress,groupedaddress]{revtex4-1}
\usepackage{graphicx}    
\usepackage{dcolumn}    
\usepackage{bm}             
\usepackage{amssymb}   
\usepackage{amsmath}    
\usepackage{mathrsfs}    
\usepackage{commath}   
\usepackage{subfigure}    
\usepackage{braket}
\usepackage{natbib}
\usepackage{float}
\usepackage[titletoc,title]{appendix}
\begin{document}

\title{Shot noise dominant regime for ellipsoidal nanoparticles in a linearly polarized beam}
\author{Changchun Zhong$^1$}
\email{zchangch@purdue.edu}
\author{F. Robicheaux$^{1,2}$}
\email{robichf@purdue.edu} 
\affiliation{$^1$Department of Physics and Astronomy, Purdue University, West Lafayette IN, 47907 USA}\affiliation{ $^2$Purdue Quantum Center, Purdue University, West Lafayette, IN, 47907, USA}

\date{\today}

\begin{abstract}

Results on the heating and the parametric feedback cooling of an optically trapped anisotropic nanoparticle in the laser shot noise dominant regime are presented. The related dynamical parameters, such as the oscillating frequency and shot noise heating rate, depend on the shape of the trapped particle. For an ellipsoidal particle, the ratio of the axis lengths and the overall size controls the shot noise heating rate relative to the frequency. For a particle with smaller ellipticity or bigger size, the relative heating rate for rotation tends to be smaller than that for translation indicating a better rotational cooling. For one feedback scheme, we also present results on the lowest occupation number that can be achieved as a function of the heating rate and the amount of classical uncertainty in the position measurement.

\end{abstract}

\maketitle

\section{introduction}

The transition between a quantum and a classical description of a system as its size is increased has been discussed extensively since the birth of quantum mechanics \cite{DATA,DTMP,DATT,CDRE}. Understanding the behavior of increasingly large systems in terms of quantum mechanics is one of the motivations for investigating mesoscopic quantum phenomena \cite{COUA,QSOM}. In order to observe mesoscopic quantum coherence, a mesoscopic system needs to be cooled to the quantum regime and it should be well isolated from its environment such that the quantum coherence is not destroyed before any observation. Recently, laser levitated nanoparticles have become a promising candidate to study mesoscopic quantum phenomena due to this system's favorable properties regarding decoherence and thermalization.\cite{COUA,LCOA,QGSA,SFDU,TOOA}.  

Despite the great advantage of laser levitation, the nanoparticle still suffers from shot noise due to photon scattering from the trapping laser. In ultrahigh vacuum, this shot noise is the dominant source of decoherence \cite{DMOP}, which will lead to an increase in energy of the solid-body degrees of freedom: the center of mass motion and the solid-body rotations. Thus, in a laser levitated cooling experiment, the photon scattering, as an unavoidable factor, plays the role of setting a fundamental cooling limit to the system since the heating from shot noise will counteract whatever method is used to cool the nanoparticle.

Cooling and controlling the center of mass vibration of levitated nanoparticles have been discussed intensively in the past several years \cite{SPFC,NMCA,NWOL,NSSO}. The interest in the rotational motion of a non-spherical nanoparticle is also increasing \cite{DORD,SDON,TOOA,FRCO}. The anisotropy of a dielectric nanoparticle has an orientation dependent interaction with a linearly polarized optical field which leads to a restricted, librational motion in some of the orientation angles when the laser intensity is large enough \cite{CAMO,RTCC}. The oscillating frequency of the rotational degrees of freedom can be much larger than that of the spatial degrees of freedom indicating that the rotational ground state can be reached at a higher temperature \cite{TOOA}. However, this feature does not guarantee that the ground state of the librational motion is easier to reach than that for the center of mass vibration. From our previous study \cite{DORD}, the decoherence rate due to shot noise in the rotational degrees of freedom was several orders of magnitude faster than that in the translational degrees of freedom for a nanoparticle interacting with blackbody radiation. The results from this theoretical study suggested that cooling the center of mass vibrations has a practical advantage over cooling the librational motion.

In this paper, we investigate the shot noise heating and parametric feedback cooling \cite{SPFC} of a nano ellipsoid trapped in a linearly polarized laser beam. The nanoparticle is trapped in the center of the beam with its long axis closely aligned with the laser polarization direction. Because the nanoparticle is nearly oriented with the laser polarization, the decoherence and shot noise heating rate of the librational motion is qualitatively changed from that for a nanoparticle interacting with blackbody radiation. The heating rate differs in the rotational and the translational degrees of freedom depending on the particle size and geometry. Importantly, we find that the relative rotational heating rate is slower than translation for a wide range of nanoparticle sizes and shapes, suggesting a better rotational than translational cooling. \textit{However}, the preference for smaller relative heating rates becomes much less certain when classical feedback uncertainty is included in the calculation. By one measure, a lower optimal cooling limit can be reached for motions with a higher relative heating rate. Thus, the details of the limitations imposed by the classical measurement uncertainty will determine whether lower quantum numbers can be achieved for vibrations or librations. The results of the feedback cooling calculations are suggestive, instead of definitive, because they are based on classical mechanics. Quantum calculations with more realistic measurement assumption would allow for estimates of the feedback cooling limits \cite{QMOC,QMAC,ASIT,FCOQ,CRAT}. Although more computationally demanding, a quantum version of feedback cooling of levitated nanoparticles should be within reach. 

This paper is organized as follows. Section \ref{s2} introduces the translational and rotational shot noise heating of a nano-ellipsoid trapped in a laser beam based on the theory of collisional decoherence \cite{CDRE,DORD,SDON}. Section \ref{s3} analyzes the particle's vibrational and librational motions and discusses its relative cooling in the laser beam. Section \ref{s4} presents the numerical results of the heating and the parametric feedback cooling. The simulation is classical and assumes an ideal measurement of the particle's position and velocity. Although limited to the classical regime, the calculations give insight into the relative difficulty of cooling the vibrational and librational degrees of freedom. In Sec. \ref{s5}, the results of feedback cooling with classical feedback uncertainty are presented. Finally, Sec. \ref{s6} summarizes our results.

\begin{figure}
\includegraphics[width=6.0cm,height=4.0cm]{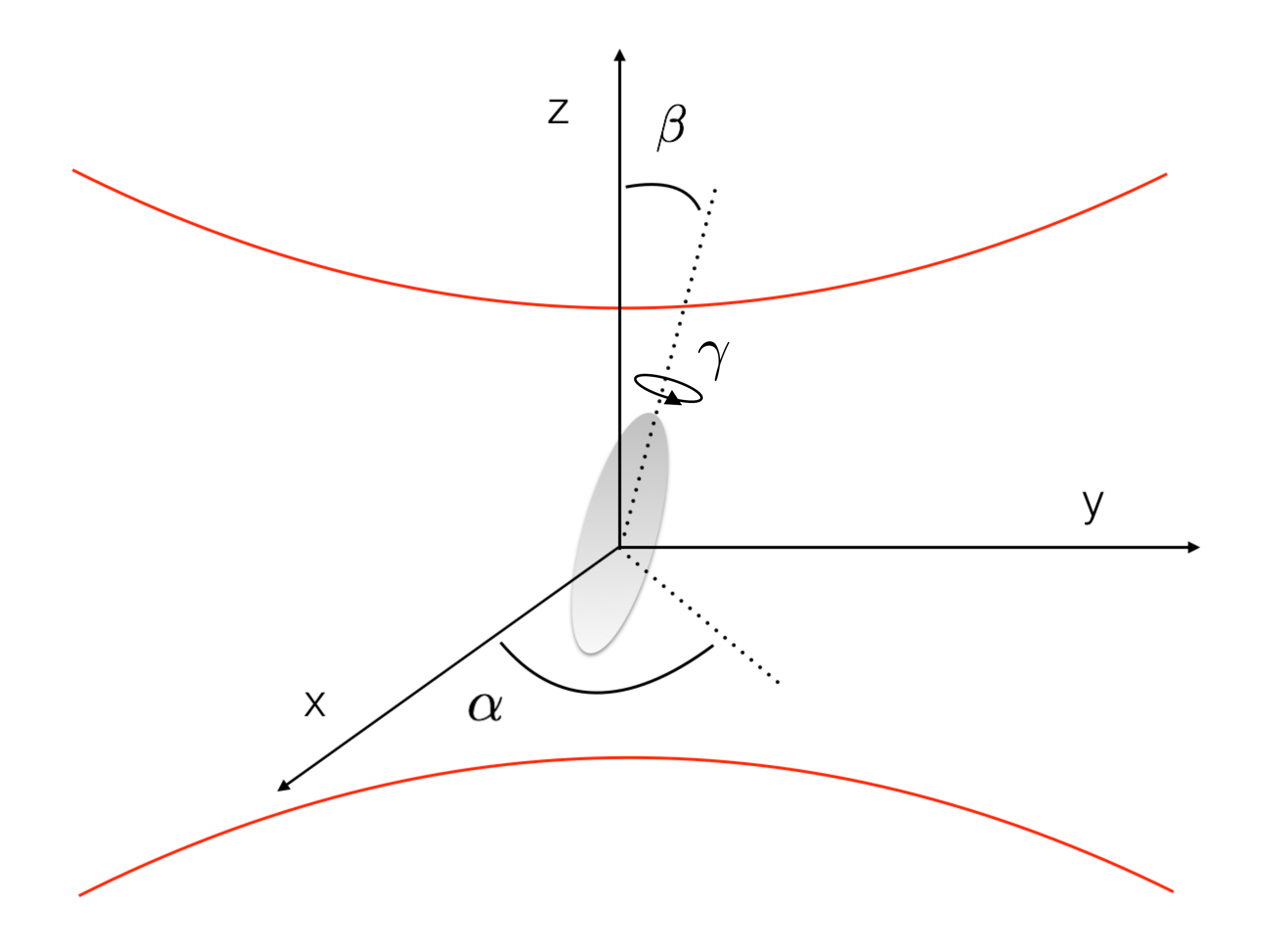}
\caption{A symmetric ellipsoidal nanoparticle is trapped in a laser beam (shown by the red line), which is polarized in the $z$ direction and propagating in the positive $y$ direction (shown by the red arrow). Besides the vibrational motion in the center of mass degrees of freedom, the ellipsoid also rotationally vibrates with its long axis closely aligned with the laser polarization direction. The angles $\alpha,\beta,\gamma$ denote an orientation of the nanoparticle. \label{fg1}}
\end{figure}

\section{Shot noise heating in a laser beam} \label{s2}

In this paper, we consider a nano ellipsoid with a size about $50\text{ nm}$ and mass $m$ trapped in a linearly polarized laser beam, as shown in Fig. (\ref{fg1}). The laser field is polarized in $z$ and propagating in the positive $y$ direction, which can be denoted by $\vec{E}_{inc}=\vec{\xi}E\exp(i\vec{k}_0\cdot\vec{r})$, where $\vec{\xi}$, $E$ and $\vec{k}_0=k_0\hat{y}$ are the polarization vector, the field magnitude and wave number respectively. The system is assumed to be well isolated from its environment and recoil from the elastically scattered photons is the major source of decoherence. 

\subsection{Shot noise in translational degrees of freedom}

In order to compare the shot noise in rotation and translation, we first present the well known photon recoil heating of a trapped nanoparticle in its center of mass motion \cite{COUA,DMOP}. Classically, the levitated nanoparticle experiences a momentum kick from each scattered photon \cite{DMOP}, each of which gives a recoil energy $\Delta E=\hbar^2k^2/2m$ when the nanoparticle is much smaller than the wavelength of the light. The shot noise heating rate can be derived through multiplying the recoil energy by the momentum transfer cross section and the photon flux. Quantum mechanically, the interaction between the system and the incoming photons causes a decoherence in the system state \cite{DATA}, which generates a diffusion in momentum space. The classical and quantum mechanical treatments lead to the same shot noise heating rate. In the position basis, the master equation can be written as \cite{CDRE}
\begin{equation}
\label{e1}
\frac{\partial}{\partial t}\rho(x,x^\prime)=-\Lambda (x,x')\rho(x,x^\prime).
\end{equation}
The unitary part of the time evolution is not shown in the above expression. $\Lambda (x,x')$ is the decoherence rate. In a long-wavelength approximation (which is a good approximation in the cases we consider), the decoherence rate $\Lambda=\mathscr{D}(x-x^{\prime})^2$, where $\mathscr{D}$ is the momentum diffusion constant and it takes the form
\begin{equation}
\label{e2}
\mathscr{D}=J_p\int d^3\vec{k}\mu(\vec{k})\int{d^2\hat{k}^\prime}\abs{f(\vec{k},\vec{k}^\prime)}^2\frac{k^2}{2}\abs{\hat{k}-\hat{k}^\prime}^2,
\end{equation}
where $J_p$ is the photon flux, $\mu(\vec{k})$ is the incoming wave number distribution and $d\sigma/d\Omega=|f(\vec{k},\vec{k}^\prime)|^2$ is the differential cross section. $\vec{k}$ and $\vec{k}^\prime$ are the incoming and outgoing wave vectors, respectively. The shot noise heating rate can be evaluated by the following formula
\begin{equation}
\label{e3}
\dot{E}_T=\frac{d\braket{\textbf{H}_T}}{dt}=\text{tr}(\textbf{K}_T\frac{\partial}{\partial t}\bm{\rho}),
\end{equation}
where $\textbf{H}_T=\textbf{K}_T+\textbf{V}_T$ denotes the system Hamiltonian and $\textbf{K}_T=\textbf{P}^2/2m$ is the free system Hamiltonian. The potential energy $\textbf{V}_T$ is absent from the right hand side of Eq. (\ref{e3}) since the trace operation will set $\Lambda=\mathscr{D}(x-x^{\prime})^2$ to zero. Combining the above equations, a straightforward calculation yields the following result
\begin{equation}
\label{e4}
\dot{E}_T=J_p\int d^3\vec{k}\mu(\vec{k})\int{d^2\hat{k}^\prime}\frac{d\sigma}{d\Omega}\frac{\hbar^2k^2}{2m}2(1-\cos\theta),
\end{equation}
where $\theta$ is the angle between the incoming and outgoing wave vector. Equation (\ref{e4}) gives the translational shot noise heating rate, which is exactly the same as what one would expect from a classical derivation \cite{SPFC}.

In order to compare the above calculation with experimental results \cite{DMOP}, Eq. (\ref{e4}) needs to be further evaluated. We are interested in the shot noise of a system coherently illuminated by a laser beam, so the incoming wave vector distribution can be approximated by
\begin{equation}
\label{e5}
\mu({\vec{k}})=\delta(\vec{k}-\vec{k}_0),
\end{equation}
in which $\vec{k}_0=k_0\hat{y}$ is the incoming wave vector. If we denote $\xi^\prime$ as the polarization vector of the outgoing wave, the scattering amplitude can be written as \cite{CEJD}
\begin{equation}
\label{e6}
f(\vec{k}^\prime,\vec{k})=\frac{k^2}{4\pi\epsilon_0E}\vec{\xi}^\prime\cdot\vec{P},
\end{equation}
where $\vec{P}=\alpha\cdot\vec{E}_{inc}$ is the induced dipole moment. For now, we choose a spherical nanoparticle (a non-spherical particle is discussed below), such that the polarizability is a scalar
\begin{equation}
\alpha=4\pi\epsilon_0\left(\frac{\epsilon-1}{\epsilon+2}\right)r^3,
\end{equation}
where $r$ is the radius, $\epsilon$ and $\epsilon_0$ are the relative and the vacuum dielectric constant respectively. Substituting the above equations into Eq. (\ref{e4}) and using the following formula \cite{QOSZ}
\begin{equation}
\label{e8}
\sum_{\lambda=(1,2)}\epsilon_i^\lambda\epsilon_j^\lambda=\delta_{ij}-\hat{k}_i\hat{k}_j
\end{equation}
to average the polarization of the outgoing wave, the shot noise heating rate is obtained
\begin{equation}
\label{e9}
\dot{E}_T=\mathscr{D}\frac{\hbar^2}{m}=\frac{8\pi J_p}{3}\left(\frac{k^2_0}{4\pi\epsilon_0}\right)^2\alpha^2\frac{\hbar^2k^2_0}{2m}.
\end{equation}

Using the parameters in Ref. \cite{DMOP}, the laser wavelength $\lambda=1064\text{ nm}$, the particle mass of a fused silica of radius $r=50\text{ nm}$ is approximately $1.2\times10^{-18}\text{ kg}$, the relative dielectric constant is about $2.1$, and the photon flux $J_p$ is equal to the laser intensity over the energy of a photon. The laser intensity at the focus is given by $I=Pk^2NA^2/2\pi$. The laser power is $P=70\text{ mW}$ and $NA=0.9$ is the numerical aperture for focusing \cite{DMOP} (These values are used throughout the paper unless specified otherwise). Combining all of these factors, the translational shot noise heating rate is
\begin{equation}
\dot{E}_T\simeq200\text{ mK/sec},
\end{equation}
which matches well the experimental result in Ref. \cite{DMOP}.

\subsection{Shot noise in rotational degrees of freedom}

Inspired by the experiment of laser trapping and cooling of non-spherical nanoparticles \cite{TOOA,FRCO}, the master equation of rotational decoherence was studied for either mass particles or thermal photons scattered from an anisotropic system, and a squared sine dependence on the orientation difference was found in the angular localization rate \cite{DORD,SDON}. Similar to the momentum diffusion induced by the translational decoherence, the rotational decoherence generates an angular momentum diffusion, which was discussed for a spherically symmetric environment in Ref. \cite{SDON}. Based on the rotational master equation, the time evolution of the expectation value of the angular momentum $\textbf{J}$ was shown to be a constant, while the second moment of the angular momentum indeed follows the diffusion equation
\begin{equation}
\braket{\textbf{J}^2}_t=\braket{\textbf{J}^2}_0+4Dt,
\end{equation}
where $D$ is the diffusion coefficient determined by different types of scattering. The diffusion coefficients of Rayleigh-type and Van der Waals-type scattering were given in Ref. \cite{SDON}. 

In this section, we discuss the rotational shot noise from photon scattering in a laser beam. The starting point is the master equation of rotational decoherence. As shown in Fig. (\ref{fg1}), the configuration of the ellipsoid can be described by its Euler angles $\ket{\Omega}=\ket{\alpha,\beta,\gamma}$ \cite{DORD,SDON}. If we denote $\rho(\Omega,\Omega^\prime)$ as the density matrix of the system in the orientational basis, the time evolution follows the equation \cite{DORD}
\begin{equation}
\label{e12}
\frac{\partial}{\partial t}\rho(\Omega,\Omega^\prime)=-\Lambda(\Omega,\Omega^\prime)\rho(\Omega,\Omega^\prime),
\end{equation}
where
\begin{equation}
\label{e13}
\Lambda=\frac{J_p}{2}\int d^3\vec{k}\mu(\vec{k})\int {d^2\hat{k}^\prime}\abs{f_\Omega(\vec{k}^\prime,\vec{k})-f_{\Omega^\prime}(\vec{k}^\prime,\vec{k})}^2
\end{equation}
is the rotational decoherence rate. Detailed discussion about the equation can be found in Ref. \cite{DORD}. Similar to Eq. (\ref{e3}), the rotational shot noise heating can be obtained by evaluating
\begin{equation}
\label{e14}
\dot{E}_R=\frac{d}{dt}\braket{\textbf{H}_R}=\text{tr}(\textbf{K}_R\frac{\partial}{\partial t}\bm{\rho}),
\end{equation}
where $\textbf{H}_R=\textbf{K}_R+\textbf{V}_R$ is the rotational Hamiltonian, $\textbf{V}_R$ is the potential energy which has zero contribution in the above equation, and $\textbf{K}_R$ is the free rotational part. For a symmetric top, $\textbf{K}_R$ takes the following form \cite{AMIQ}
\begin{equation}
\label{e15}
\begin{split}
\textbf{K}_R=-\frac{\hbar^2}{2I_1}(\frac{\partial^2}{\partial\beta^2}+\cot\beta\frac{\partial}{\partial\beta}+(\frac{I_1}{I_3}+\cot^2\beta)\frac{\partial^2}{\partial\gamma^2}\\+\frac{1}{\sin^2\beta}\frac{\partial^2}{\partial\alpha^2}-\frac{2\cos\beta}{\sin^2\beta}\frac{\partial^2}{\partial\alpha\partial\gamma} ),
\end{split}
\end{equation}
where $I_1$ and $I_3$ are the moments of inertia of the ellipsoid along the short and long axis, respectively. To calculate the shot noise heating, the next step is to determine the decoherence rate $\Lambda$. As with the derivation of the translational shot noise, the distribution of the laser wave vector takes the delta function $\mu(\vec{k})=\delta(\vec{k}-\vec{k}_0)$, where $\vec{k}_0$ is in the propagating $y$ direction. The scattering amplitude is given by 
\begin{equation}
f_\Omega(\vec{k}^\prime,\vec{k})=\frac{k^2}{4\pi\epsilon_0E}\vec{\xi}^\prime\cdot\bar{\bar{\alpha}}_\Omega\cdot\vec{E}_{inc},
\end{equation}
where $\bar{\bar{\alpha}}_\Omega$ is the polarizability matrix for a specific configuration $\ket{\Omega}=\ket{\alpha,\beta,\gamma}$. If we place the ellipsoid symmetrically along the coordinate axis, the polarizability matrix will be diagonal
\begin{equation}
\bar{\bar{\alpha}}_{0}=\left(\begin{array}{ccc}\alpha_x & 0 & 0 \\0 & \alpha_y & 0 \\0 & 0 & \alpha_z\end{array}\right),
\end{equation}
where $\alpha_x=\alpha_y$ for a symmetric top. The polarizability with other rotational configuration can be derived through the following operation 
\begin{equation}
\label{e18}
\bar{\bar{\alpha}}_{\Omega}=R^{\dag}(\Omega)\bar{\bar{\alpha}}_0 R(\Omega).
\end{equation}
Combining the above equations and averaging over the polarizations of the outgoing wave using Eq. (\ref{e8}), the integral of Eq. (\ref{e13}) becomes
\begin{equation}
\label{e19}
\begin{split}
\Lambda=\frac{J_p}{2}\frac{k_0^4}{(4\pi\epsilon_0)^2}\frac{2\pi}{3}(\alpha_z-\alpha_x)^2 ( 1-\cos(2\beta)\cos(2\beta^\prime)\\-\cos(\alpha-\alpha^\prime)\sin(2\beta)\sin(2\beta^\prime) ).
\end{split}
\end{equation}
The polarizability $\alpha_{x,z}$ should not be confused with the Euler angle $\alpha$ and $\alpha^\prime$. As expected, $\Lambda$ differs from the decoherence rate from blackbody radiation given in Ref. \cite{DORD}. The localization rate $\Lambda$ depends on the orientations $\ket{\Omega}$ and $\ket{\Omega^\prime}$ individually since the polarization of incoming photons is not isotropic. There is no dependence on $\gamma$ because we're assuming a symmetric top. The localization rate depends only on the difference of the angle $\alpha$ because the photons are linearly polarized in the z-direction which does not have a preferential angle in the xy-plane. 

For the cases considered below, we take the small oscillation approximation $\beta\ll 1$ which will be justified in the next section. (Unless specified otherwise, the symbol $\simeq$ in this paper means this approximation is used.) Combining the Eqs. (\ref{e12}), (\ref{e15}) and Eq. (\ref{e19}), a direct evaluation of Eq. (\ref{e14}) yields the rotational shot noise heating rate
\begin{equation}
\dot{E}_R\simeq\frac{8\pi J_p}{3}\left(\frac{k_0^2}{4\pi\epsilon_0}\right)^2(\alpha_z-\alpha_x)^2\frac{\hbar^2}{2I_1},
\end{equation}
where terms of order $\beta^2$ have been dropped.

\begin{table*}[htp]
\caption{The parameters for three different nano-diamonds in a laser trap. The data is ordered for diamonds with decreasing ellipticity, while their sizes $\sqrt{a^2+b^2}$ are kept approximately the same. The trapping laser has wavelength $\lambda=1064\text{ nm}$ and power $P=70\text{ mW}$. }\label{tab1}
\begin{center}
\begin{tabular}{|c|c|c|c|c|c|c|c|c|c|c|}
\hline
$(a,b)/\text{nm}$ & $\frac{\alpha_z-\alpha_x}{\alpha_z}$ & $\omega_{\beta_1}/2\pi$ & $\omega_x/2\pi$ & $\omega_y/2\pi$ & $\dot{E}_R$(mK/s) & $\dot{E}_T$(mK/s) &$ {\dot{E}_R}/{\dot{E}_T}$ & $\frac{\omega_{\beta_1}}{\omega_x}$ & $\frac{\braket{\dot{n}}_R}{\braket{\dot{n}}_T}$  & $\frac{\Delta n_R}{\Delta n_T}$  \\
\hline
$(15,70)$ & $0.60$ & $4.02$ MHz & $625$ kHz & $398$ kHz & $3.83\times10^{3}$& $ 382 $ & $10.0$ & $6.43$ & 1.56  & 0.24 \\
$(38,60)$ & $0.28$ & $2.20$ MHz & $497$ kHz & $316$ kHz & $1.84\times10^{3}$& $ 838 $ & $2.20$ & $4.42$ & 0.50 & 0.11 \\
$(48,53)$ & $0.07$ & $998$ kHz  & $454$ kHz & $289$ kHz & $113$& $ 824 $ & $0.14$ & $2.20$ & 0.06 & 0.03  \\
\hline
\end{tabular}
\end{center}
\label{default}
\end{table*}

\begin{table*}[htp]
\caption{The parameters for three different nano-diamonds in a laser trap. The data is for diamonds with increasing size while fixing the ellipticity such that the ratio $(\alpha_z-\alpha_x)/\alpha_z$ stays approximately the same. The trapping laser has wavelength $\lambda=1064\text{ nm}$ and power $P=70\text{ mW}$. }\label{tab11}
\begin{center}
\begin{tabular}{|c|c|c|c|c|c|c|c|c|c|c|}
\hline
$(a,b)/\text{nm}$ & $\frac{\alpha_z-\alpha_x}{\alpha_z}$ & $\omega_{\beta_1}/2\pi$ & $\omega_x/2\pi$ & $\omega_y/2\pi$ & $\dot{E}_R$(mK/s) & $\dot{E}_T$(mK/s) &$ {\dot{E}_R}/{\dot{E}_T}$ & $\frac{\omega_{\beta_1}}{\omega_x}$ & $\frac{\braket{\dot{n}}_R}{\braket{\dot{n}}_T}$ & $\frac{\Delta n_R}{\Delta n_T}$    \\
\hline
$(27,42)$ & $0.28$ & $3.14$ MHz & $497$ kHz & $316$ kHz & $1.23\times10^{3}$& $ 292 $ & $4.22$ & $6.31$ & 0.68 & 0.11    \\
$(38,60)$ & $0.28$ & $2.20$ MHz & $497$ kHz & $316$ kHz & $1.84\times10^{3}$& $ 838 $ & $2.20$ & $4.42$ & 0.50 & 0.11  \\
$(49,78)$ & $0.28$ & $1.68$ MHz  & $497$ kHz & $316$ kHz & $2.46\times10^{3}$& $ 1830 $ & $1.34$ & $3.40$ & 0.39 &0.11   \\
\hline
\end{tabular}
\end{center}
\label{default}
\end{table*}


\begin{table*}[htp]
\caption{The parameters for three different fused silica in a laser trap. The data is for silica with different ellipticities, while their sizes $\sqrt{a^2+b^2}$ are kept approximately the same. The trapping laser has wavelength $\lambda=1064\text{ nm}$ and power $P=70\text{ mW}$. }\label{tab2}
\begin{center}
\begin{tabular}{|c|c|c|c|c|c|c|c|c|c|c|}
\hline
$(a,b)/\text{nm}$ & $\frac{\alpha_z-\alpha_x}{\alpha_z}$ & $\omega_{\beta_1}/2\pi$ & $\omega_x/2\pi$ & $\omega_y/2\pi$ & $\dot{E}_R$(mK/s) & $\dot{E}_T$(mK/s) &$ {\dot{E}_R}/{\dot{E}_T}$ & $\frac{\omega_{\beta_1}}{\omega_x}$ & $\frac{\braket{\dot{n}}_R}{\braket{\dot{n}}_T}$  & $\frac{\Delta n_R}{\Delta n_T}$  \\
\hline
$(15,70)$ & $0.30$ & $1.90$ MHz & $419$ kHz & $267$ kHz & $119$& $ 48.6 $ & $2.45$ & $4.52$ & 0.54 & 0.12   \\
$(38,60)$ & $0.13$ & $1.17$ MHz & $388$ kHz & $247$ kHz & $93.2$& $ 197 $ & $0.47$ & $3.01$ & 0.16 & 0.05 \\
$(48,53)$ & $0.03$ & $549$ kHz  & $374$ kHz & $238$ kHz & $6.50$& $ 240 $ & $0.03$ & $1.47$ & 0.02 & 0.01   \\
\hline
\end{tabular}
\end{center}
\label{default}
\end{table*}

\begin{table*}[htp]
\caption{The parameters for three different fused silica in a laser trap. The data is for silica with increasing sizes while the ellipticity is fixed such that the ratio $(\alpha_z-\alpha_x)/\alpha_z$ stays approximately the same. The trapping laser has wavelength $\lambda=1064\text{ nm}$ and power $P=70\text{ mW}$. }\label{tab22}
\begin{center}
\begin{tabular}{|c|c|c|c|c|c|c|c|c|c|c|}
\hline
$(a,b)/\text{nm}$ & $\frac{\alpha_z-\alpha_x}{\alpha_z}$ & $\omega_{\beta_1}/2\pi$ & $\omega_x/2\pi$ & $\omega_y/2\pi$ & $\dot{E}_R$(mK/s) & $\dot{E}_T$(mK/s) &$ {\dot{E}_R}/{\dot{E}_T}$ & $\frac{\omega_{\beta_1}}{\omega_x}$ & $\frac{\braket{\dot{n}}_R}{\braket{\dot{n}}_T}$ & $\frac{\Delta n_R}{\Delta n_T}$   \\
\hline
$(27,42)$ & $0.13$ & $1.67$ MHz & $388$ kHz & $247$ kHz & $62.6$& $ 69.1 $ & $0.91$ & $4.30$ & 0.21 & 0.05   \\
$(38,60)$ & $0.13$ & $1.17$ MHz & $388$ kHz & $247$ kHz & $93.2$& $ 197 $ & $0.47$ & $3.01$ & 0.16 & 0.05  \\
$(49,78)$ & $0.13$ & $899$ kHz  & $388$ kHz & $247$ kHz & $124$& $ 427 $ & $0.29$ & $2.31$  & 0.12  &0.05 \\
\hline
\end{tabular}
\end{center}
\label{default}
\end{table*}


\section{Relative cooling of the ellipsoid in the laser beam} \label{s3}

There are several possible quantities that are useful when comparing the cooling of translation and rotation. The first one is the ratio of magnitudes of the translational and rotational shot noise, which is written as
\begin{equation}
\label{e22}
\frac{\dot{E}_R}{\dot{E}_T}\simeq 5\left(\frac{\lambda}{2\pi\sqrt{a^2+b^2}}\right)^2\frac{(\alpha_z-\alpha_x)^2}{\alpha_z^2},
\end{equation}
where the moment of inertia $I_1=\frac{1}{5}m(a^2+b^2)$ with $a$ and $b$ being the short and long axis of the ellipsoid and $k_0=2\pi/\lambda$ are used. The polarizability can be determined by the formula \cite{AASO}
\begin{equation}
\alpha_i=\epsilon_0V\frac{\epsilon-1}{1+L_i(\epsilon-1)},
\end{equation}
where $V$ is the particle volume, and $\epsilon$ is the relative dielectric constant. $L_{i=(x,y,z)}$ is determined by
\begin{equation}
\begin{split}
L_x&=L_y=\frac{1-L_z}{2},\\
L_z&=\frac{1-e^2}{e^2}(\frac{1}{2e}\ln\frac{1+e}{1-e}),
\end{split}
\end{equation}
where $e=\sqrt{1-a^2/b^2}$ is the ellipticity of the nanoparticle. Using the wavelength $\lambda=1064\text{ nm}$ and $\epsilon=5.7$ for diamonds and $\epsilon=2.1$ for silica, the rotational and translational shot noise and their ratios ${\dot{E}_R}/{\dot{E}_T}$ for several nano-diamonds and fused silica are given in Tab. \ref{tab1}, \ref{tab11}, \ref{tab2} and \ref{tab22} (For convenience, other related quantities are included in the tables). The geometries of the ellipsoids in the tables are chosen in a way such that their sizes $\sqrt{a^2+b^2}$ or ellipticities are approximately fixed. From the table, we see that the ratio ${\dot{E}_R}/{\dot{E}_T}$ differs depending on the ellipticity or size of the nanoparticle. More elongated or smaller ellipsoid tends to have higher shot noise heating in the rotational degrees of freedom, which suggests that particles with more spherical shape or bigger size may be better for rotational cooling.

The second useful quantity is the ratio of the rate of change of occupation number $\braket{\dot{n}}_R/\braket{\dot{n}}_T$, where $\braket{n}\equiv E/\hbar\omega$ is defined as the mean occupation number, and $E$ and $\omega$ are the energy and the oscillating frequency in the corresponding degree of freedom. For exploration of quantum phenomena, the occupation should be as small as possible. In order to get the ratio, it is necessary to analyze the mechanical motion of the nanoparticle in the laser trap. We consider an incident Gaussian beam which is $z$ polarized and propagates in the $y$ direction, as shown in Fig. (\ref{fg1}). The detailed discussion of the Gaussian beam can be found in Ref. \cite{OAAC,SPFC}. The ellipsoid in the laser trap experiences a force and a torque
\begin{equation}
\begin{split}
F_i&=\frac{1}{2}(\vec{P}\cdot\partial_i\vec{E}_{inc}),\\
M_i&=\frac{1}{2}(\vec{P}\times\vec{E}_{inc})_i,
\end{split}
\end{equation}
where no absorption is assumed such that the dipole moment $\vec{P}$ is real. For the center of mass motion, using the small oscillation approximation, the particle oscillates harmonically in the trap and each degree of freedom has an oscillating frequency,
\begin{equation}
\begin{split}
\omega_x=\omega_z&\simeq\sqrt{\frac{\alpha_z}{m}}\frac{E_0}{w_0},\\
\omega_y&\simeq\sqrt{\frac{\alpha_z}{2m}}\frac{E_0}{y_0},
\end{split}
\end{equation}
where all corrections quadratic in the amplitude of oscillations have been dropped. $y_0={\pi w_0^2}/{\lambda}$, $w_0=\lambda/(\pi NA)$ is the beam waist and $E_0$ is the field strength in the center of the laser focus.  Similarly, for the rotational motion, due to the torque exerted on the particle, the long axis of the ellipsoid will be aligned with the direction of the laser polarization, as shown in Fig. ({\ref{fg1}). From the small oscillation approximation, the torsional oscillating frequencies can be written as 
\begin{equation}
\begin{split}
\omega_{\beta_1}=\omega_{\beta_2}\simeq\sqrt{\frac{\alpha_z-\alpha_x}{2I_1}}E_0,
\end{split}
\end{equation}
where all corrections quadratic in the amplitude of oscillations have been dropped. The subindex $\beta_1$ and $\beta_2$ are used to denote the torsional vibration along the $x$ and $y$ axis, respectively. From the above equations, one finds that the ratio of torsional oscillating frequency to the translational oscillating frequency is aproximately given by
\begin{equation}
\label{e26}
\frac{\omega_{\beta_1}}{\omega_x}\simeq\frac{\sqrt{5}w_0}{\sqrt{2(a^2+b^2)}}\sqrt{\frac{\alpha_z-\alpha_x}{\alpha_z}}.
\end{equation}
In an experiment, the beam waist is much bigger than the size of the particle, and the polarizability $\alpha_z$ and $\alpha_z-\alpha_x$ are roughly the same order, so the rotational oscillating frequency is generally higher than the translational oscillating frequency \cite{TOOA}. 

\begin{figure}
\centering
\subfigure
{
\put(23,-12){(a)}
\begin{minipage}[]{0.24\textwidth}
\includegraphics[width=3.9cm,height=3.2cm]{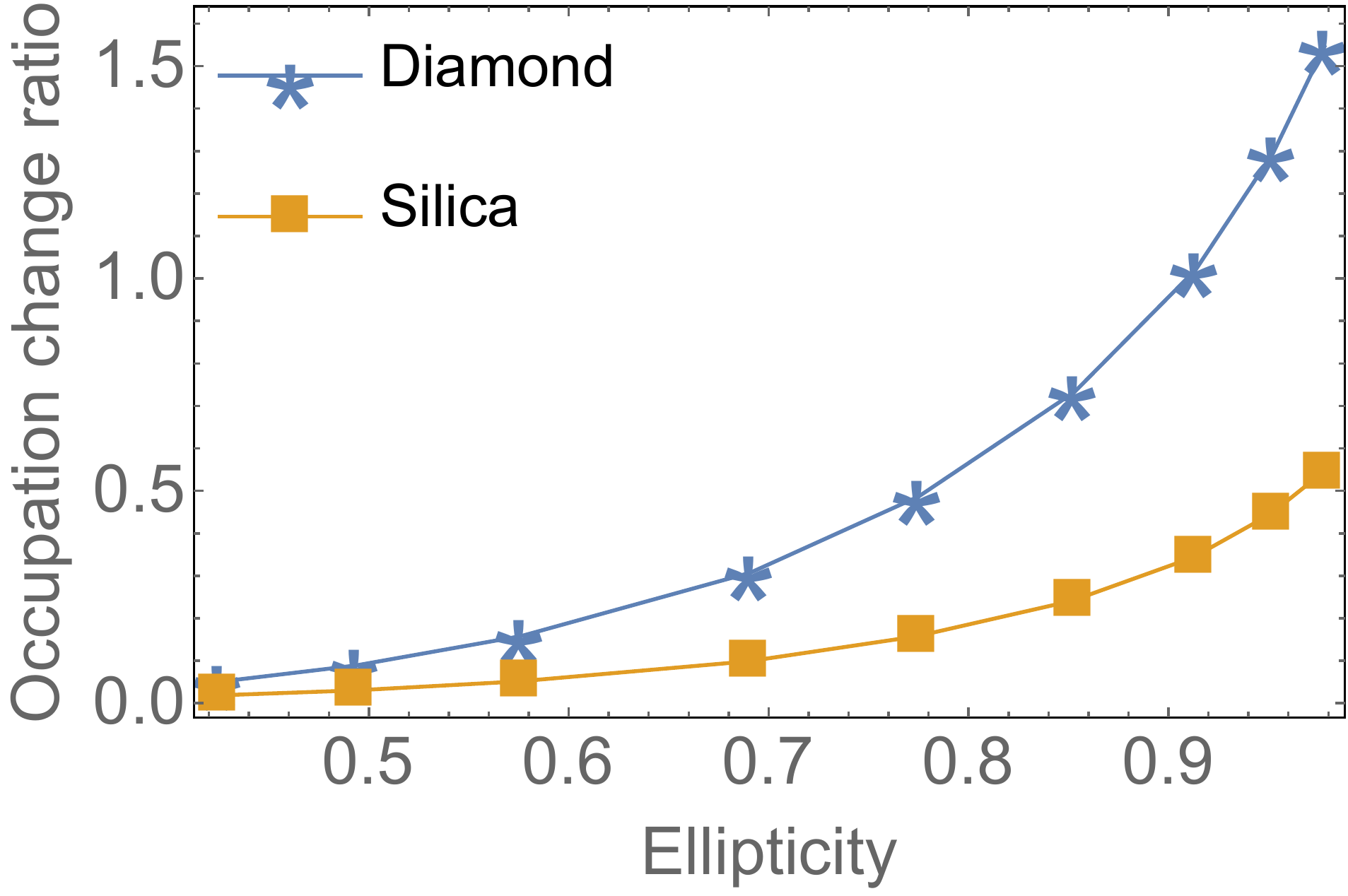}
\end{minipage}
\label{fadd1}
}
\subfigure
{
\put(22,-21){(b)}
\begin{minipage}[]{0.22\textwidth}
\includegraphics[width=3.9cm,height=3.2cm]{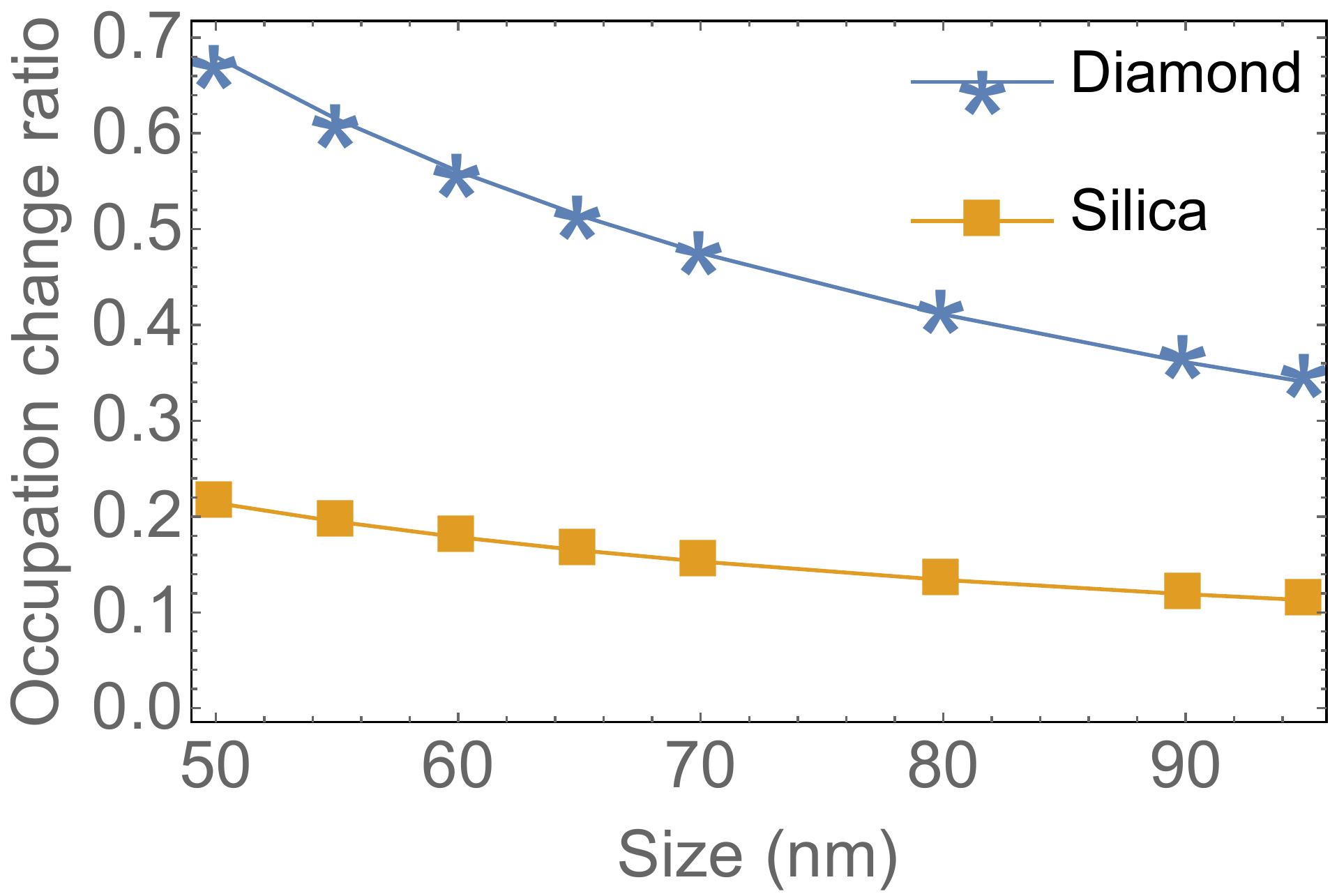}
\end{minipage}
\label{fadd2}
}
\caption{{(Color online)}  The ratio of the occupation number change $\braket{\dot{n}_R}/\braket{\dot{n}_T}$ in terms of ellipticity (\textbf{a}) and size (\textbf{b}). (\textbf{a}) The size of particles is fixed at $\sqrt{a^2+b^2}=71\text{ nm}$ while the ellipticity increases. (\textbf{b}) The ellipticity is fixed at $e=0.77$ while the particle size increases. The blue curves are for diamonds while the yellow curves are for silica. \label{fga1} } 
\end{figure}

Thus, the ratio of the corresponding rate of change of occupation number is obtained
\begin{equation}
\label{e29}
\frac{\braket{\dot{n}}_R}{\braket{\dot{n}}_T}\equiv \frac{\dot{E}_R/\omega_{\beta_1}}{\dot{E}_T/\omega_x}\simeq\frac{\lambda^2}{4\pi^2 w_0}\sqrt{\frac{10(\alpha_z-\alpha_x)^3}{(a^2+b^2)\alpha_z^3}},
\end{equation}
where the ratio is determined by the laser parameters, the particle size and the quantity $(\alpha_z-\alpha_x)/\alpha_z$ (determined by the particle ellipticity and dielectric constant). The ratios ${\braket{\dot{n}}_R}/{\braket{\dot{n}}_T}$ with respect to the particle ellipticity and size are given in Fig. (\ref{fga1}). The blue and yellow curves are for diamonds and silica respectively. In Fig. \ref{fadd1}, the particle size is kept fixed while we increase the ellipticity. As the particle shape approaches more spherical (ellipticity decreases), the ratio ${\braket{\dot{n}}_R}/{\braket{\dot{n}}_T}$ becomes smaller.  In Fig. \ref{fadd2}, we change the particle size while the particle ellipticity stays fixed. As the particle size increases, we see ${\braket{\dot{n}}_R}/{\braket{\dot{n}}_T}$ gets smaller. In addition, comparing the results for diamond and silica with the same geometries, we see that the ratio ${\braket{\dot{n}}_R}/{\braket{\dot{n}}_T}$ is generally smaller for silica. The reason is that $(\alpha_z-\alpha_x)/\alpha_z$ in Eq. (\ref{e29}) is smaller for particles with smaller dielectric constants and silica has a smaller dielectric constant than diamond. Intuitively, the ratio ${\braket{\dot{n}}_R}/{\braket{\dot{n}}_T}$ should be chosen as small as possible so as to get a better rotational cooling to the ground state. However, we will show later that the unavoidable measurement noise quantitatively modifies the trend.

\begin{figure}
\includegraphics[width=7.0cm,height=5.0cm]{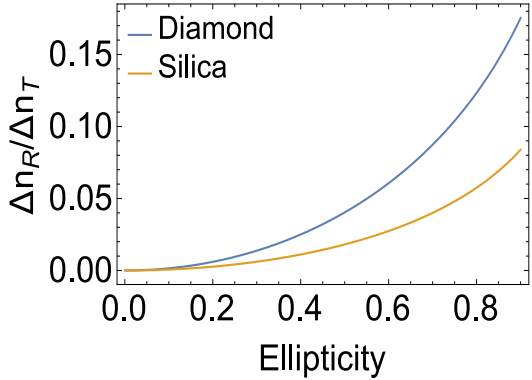}
\caption{ (Color online) The ratio $\Delta n_R/\Delta n_T$ in terms of the particle ellipticity.  The blue and yellow curves correspond to Diamond and Silica respectively.  \label{a3}}
\end{figure}

The third useful quantity is the ratio $\Delta n_R/\Delta n_T$, where $\Delta n\equiv 2\pi\dot{E}/\hbar\omega^2$ is the change in occupation number over one vibrational period in the corresponding degree of freedom. The ratio can be written as
\begin{equation}
\frac{\Delta n_R}{\Delta n_T}=\frac{\dot{E}_R/\omega_\beta^2}{\dot{E}_T/\omega_x^2}\simeq\frac{\lambda^2}{2\pi^2 w_0^2} \frac{\alpha_z-\alpha_x}{\alpha_z},
\end{equation}
which only depends on the laser parameters, the particle ellipticity and the particle dielectric constant. The ratios $\Delta n_R/\Delta n_T$ for diamond and silica with respect to the particle ellipticity are given in the tables and are plotted in Fig. (\ref{a3}). The curves show that the ratio increases with the particle ellipticity and also increases with the particle dielectric constant. This quantity is important and we will show in Sec. \ref{s5} that this quantity actually controls the classical dynamics during the feedback cooling.

\begin{figure}
\centering
\subfigure
{
\put(24,36){(a)}
\begin{minipage}[]{0.22\textwidth}
\includegraphics[width=3.9cm,height=3.2cm]{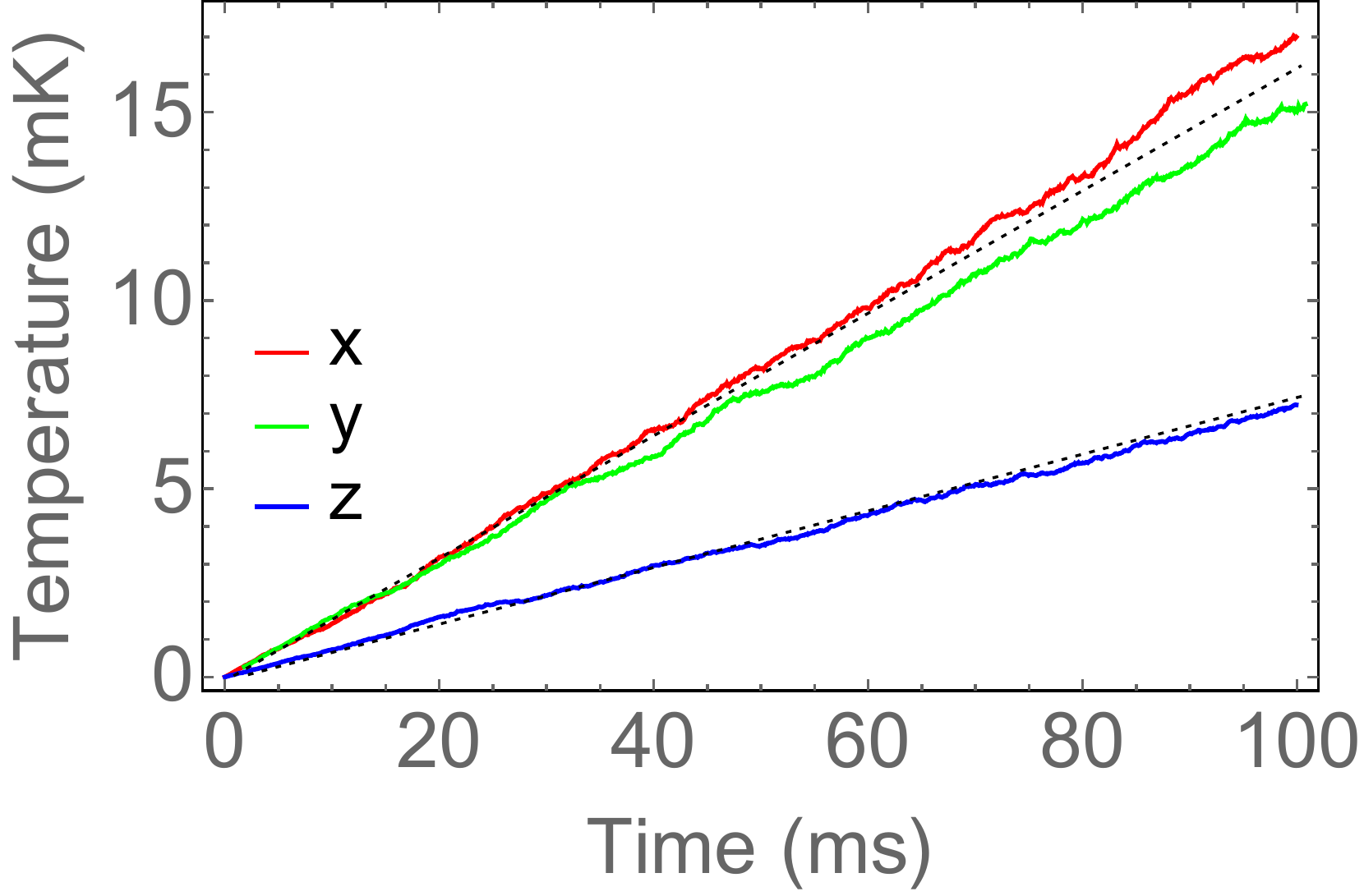}
\end{minipage}
\label{fg2a}
}
\subfigure
{
\put(24,36){(b)}
\begin{minipage}[]{0.22\textwidth}
\includegraphics[width=3.9cm,height=3.2cm]{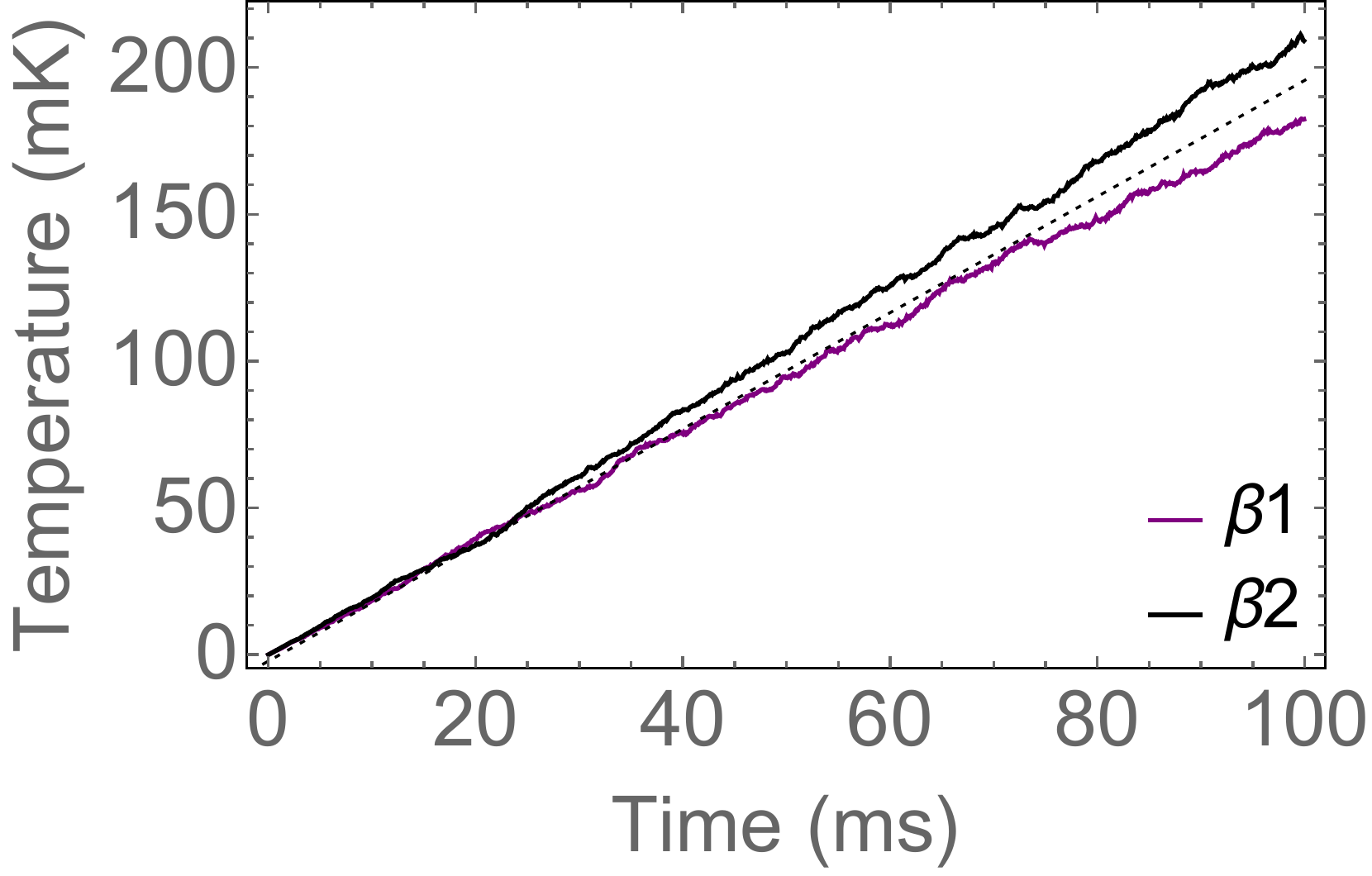}
\end{minipage}
\label{fg2b}
}
\subfigure
{
\put(20,36){(c)}
\begin{minipage}[]{0.22\textwidth}
\includegraphics[width=3.9cm,height=3.2cm]{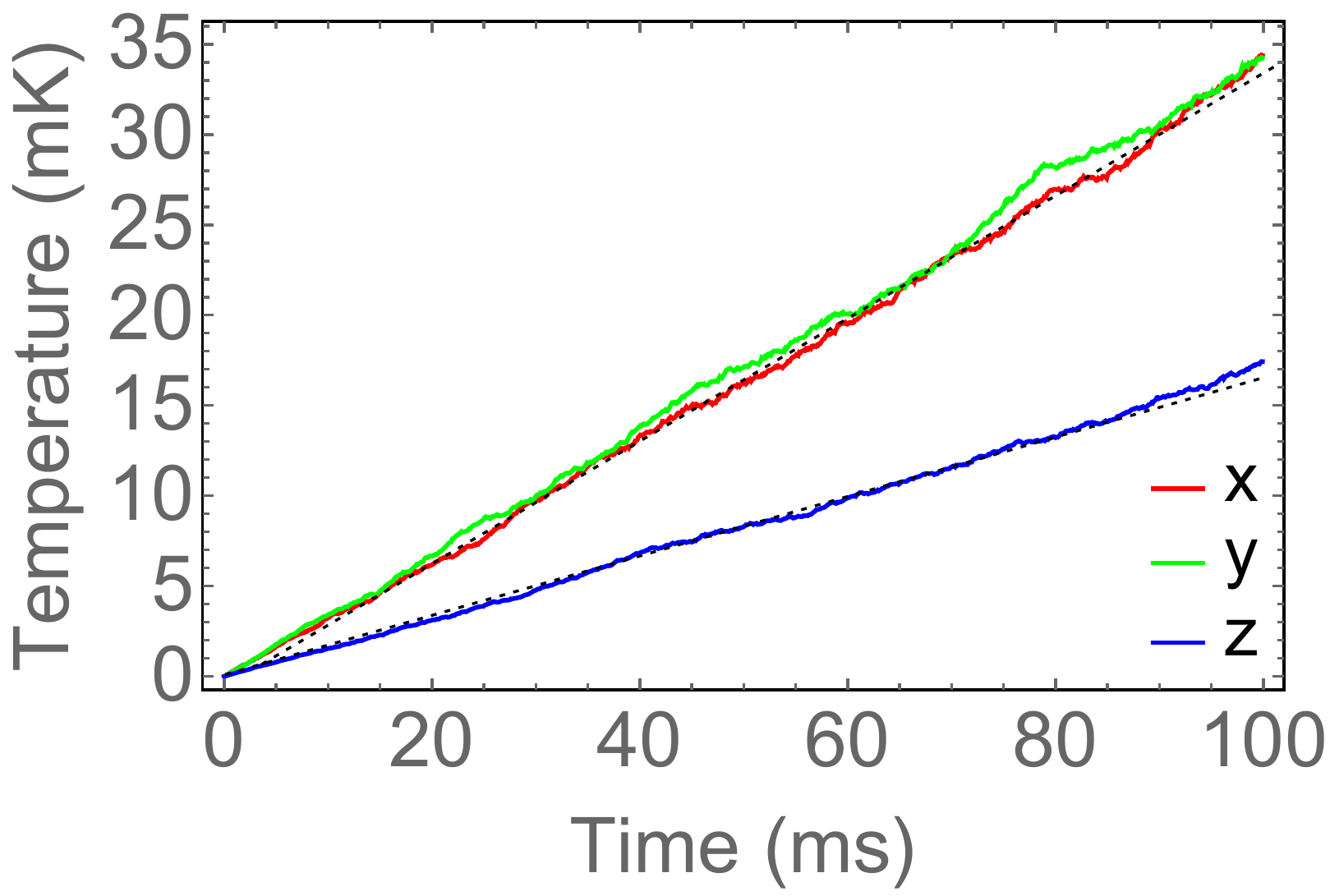}
\end{minipage}
\label{fg2c}
}
\subfigure
{
\put(23,34){(d)}
\begin{minipage}[]{0.22\textwidth}
\includegraphics[width=3.9cm,height=3.2cm]{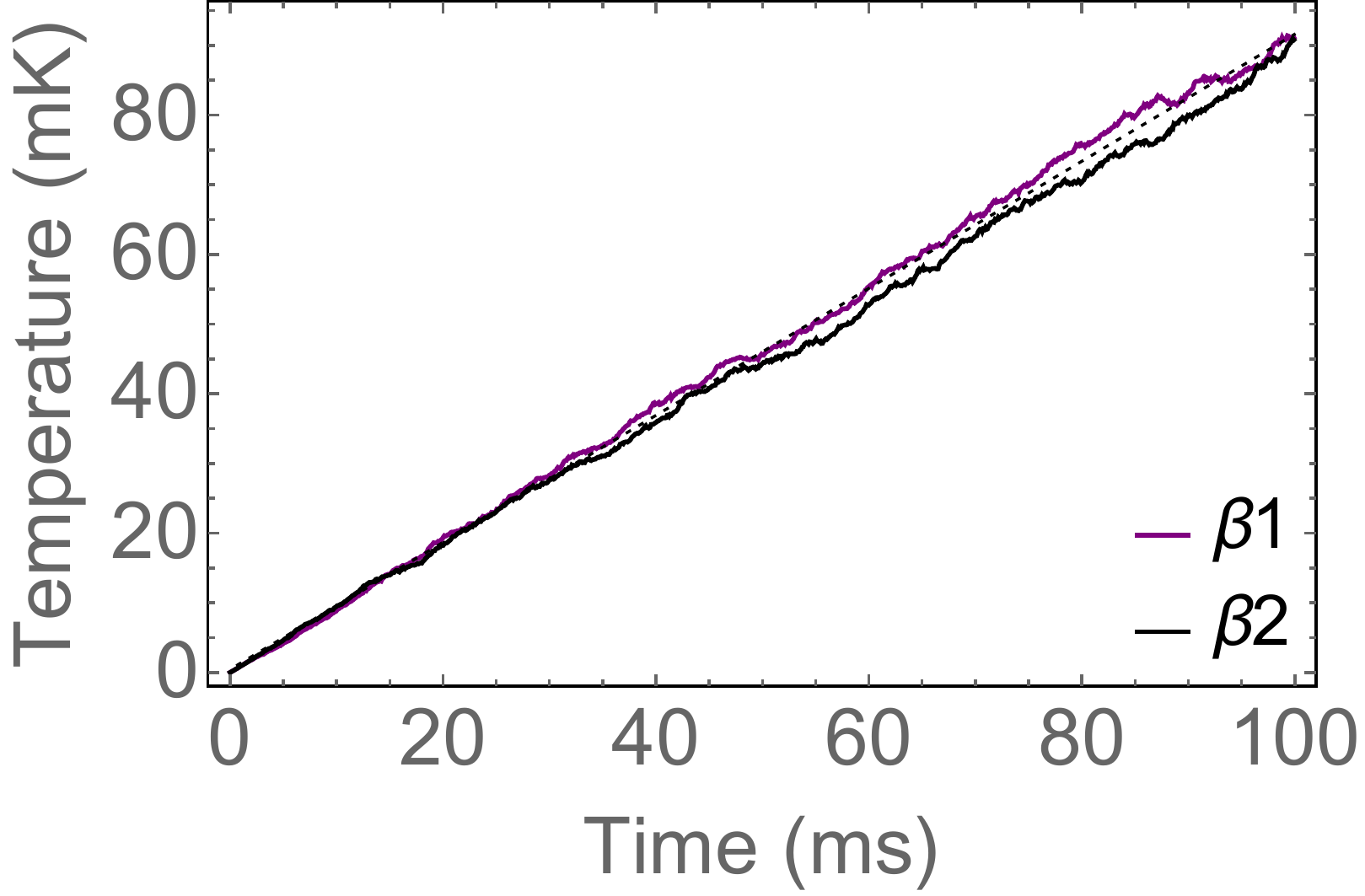}
\end{minipage}
\label{fg2d}
}

\subfigure
{
\put(20,34){(e)}
\begin{minipage}[]{0.22\textwidth}
\includegraphics[width=3.9cm,height=3.2cm]{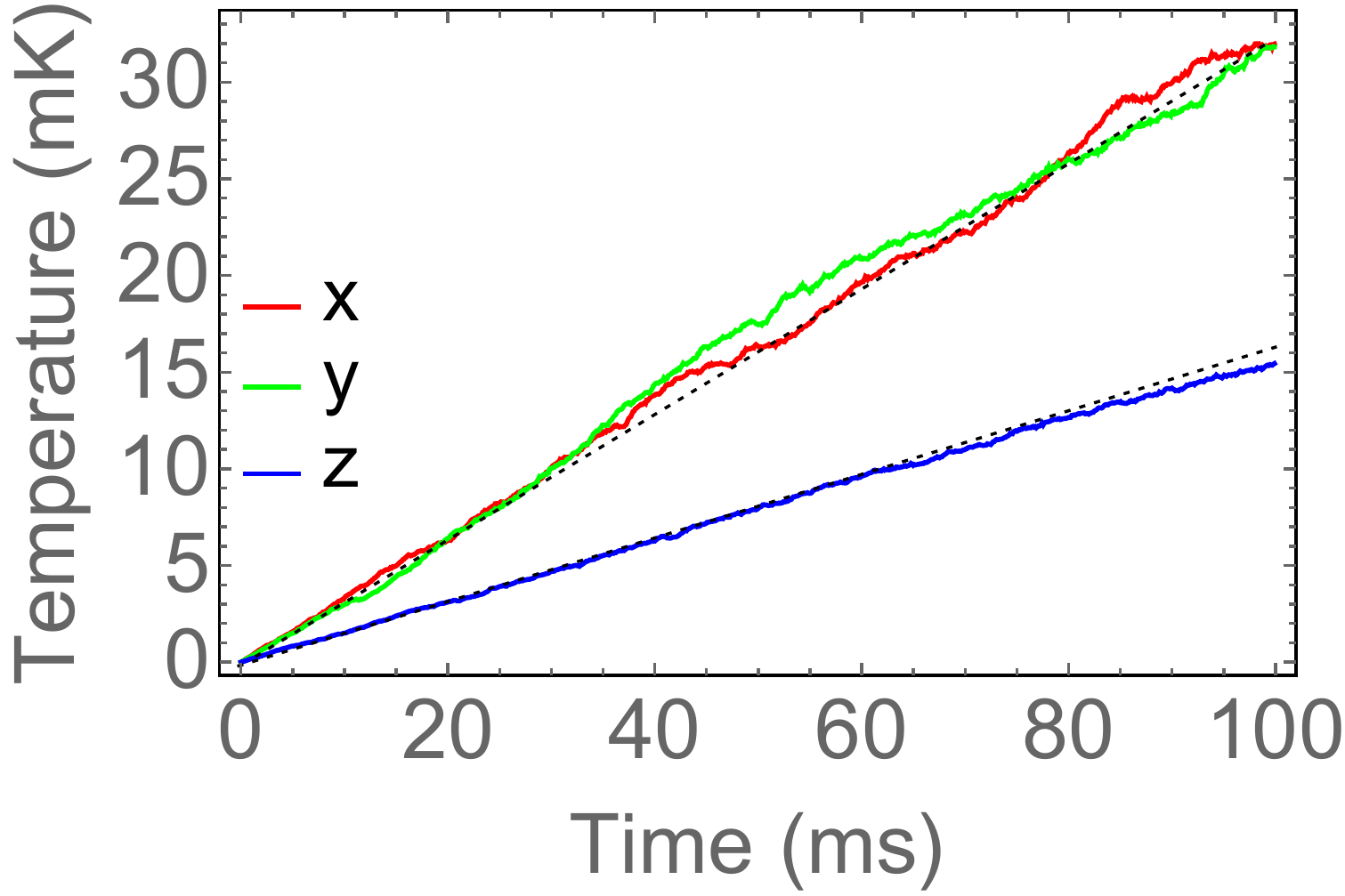}
\end{minipage}
\label{fg2e}
}
\subfigure
{
\put(24,34){(f)}
\begin{minipage}[]{0.22\textwidth}
\includegraphics[width=3.9cm,height=3.2cm]{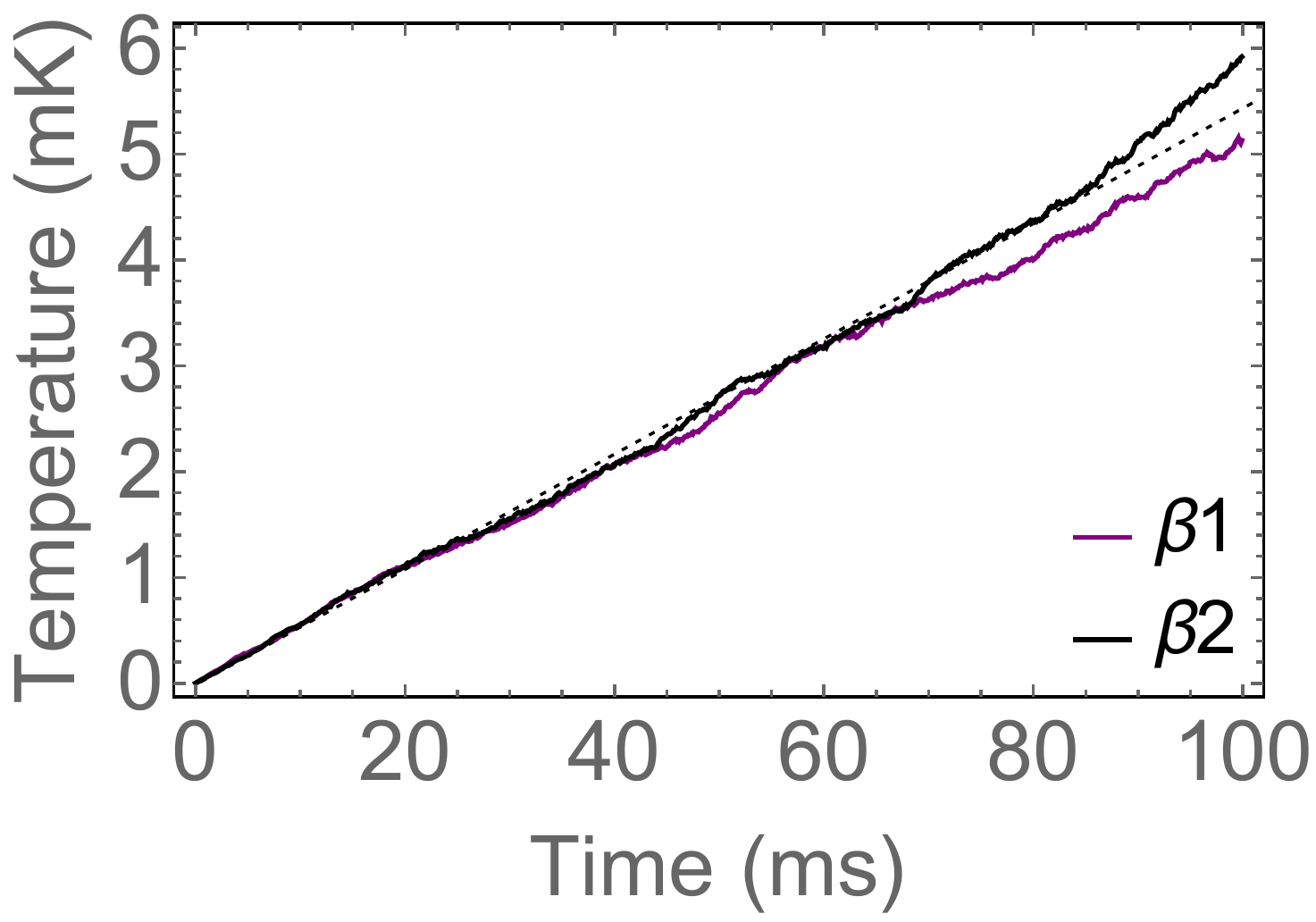}
\end{minipage}
\label{fg2f}
}

\caption{{(Color online)}  The classical simulation results of shot noise heating for nano-diamonds in both the translational and rotational degrees of freedom. Each curve is averaged over $400$ individual reheating trajectories. \textbf{(a)} and \textbf{(b)} are for the nanoparticle with half axes $(a=15\text{ nm},b=70\text{ nm})$, while \textbf{(c)} and \textbf{(d)} with half axes $(a=38\text{ nm},b=60\text{ nm})$, \textbf{(e)} and \textbf{(f)} with half axes $(a=48\text{ nm},b=53\text{ nm})$. The dashed lines are the heating curves $T=T_0+\dot{E} t$ with $T_0$ the initial temperature and $\dot{E}$ the corresponding heating rate from Tab. \ref{tab1}. \label{fg2} }
\end{figure}

The above equations are based on a small oscillation angle approximation. In a cooling experiment, the maximal oscillation angle can be estimated by 
\begin{equation}
\beta_{max}\simeq\sqrt{\frac{2k_BT}{I_1\omega_{\beta}^2}},
\end{equation}
where $k_B$ is the Boltzmann constant and $T$ denotes the temperature. Using the data ($a=48\text{ nm},b=53\text{ nm}$) from Tab. \ref{tab1} and \ref{tab2}, we find that the maximal angle spread is still small ($\beta_{max}\simeq10^{-3}$ rad for diamond, $\beta_{max}\simeq10^{-2}$ rad for silica) at $T=0.1\text{ K}$. For higher oscillating frequencies and lower temperature, the maximal angle spread $\beta_{max}$ will be even smaller.

\begin{figure}
\centering
\subfigure
{
\put(29,37){(a)}
\begin{minipage}[]{0.24\textwidth}
\includegraphics[width=3.9cm,height=3.2cm]{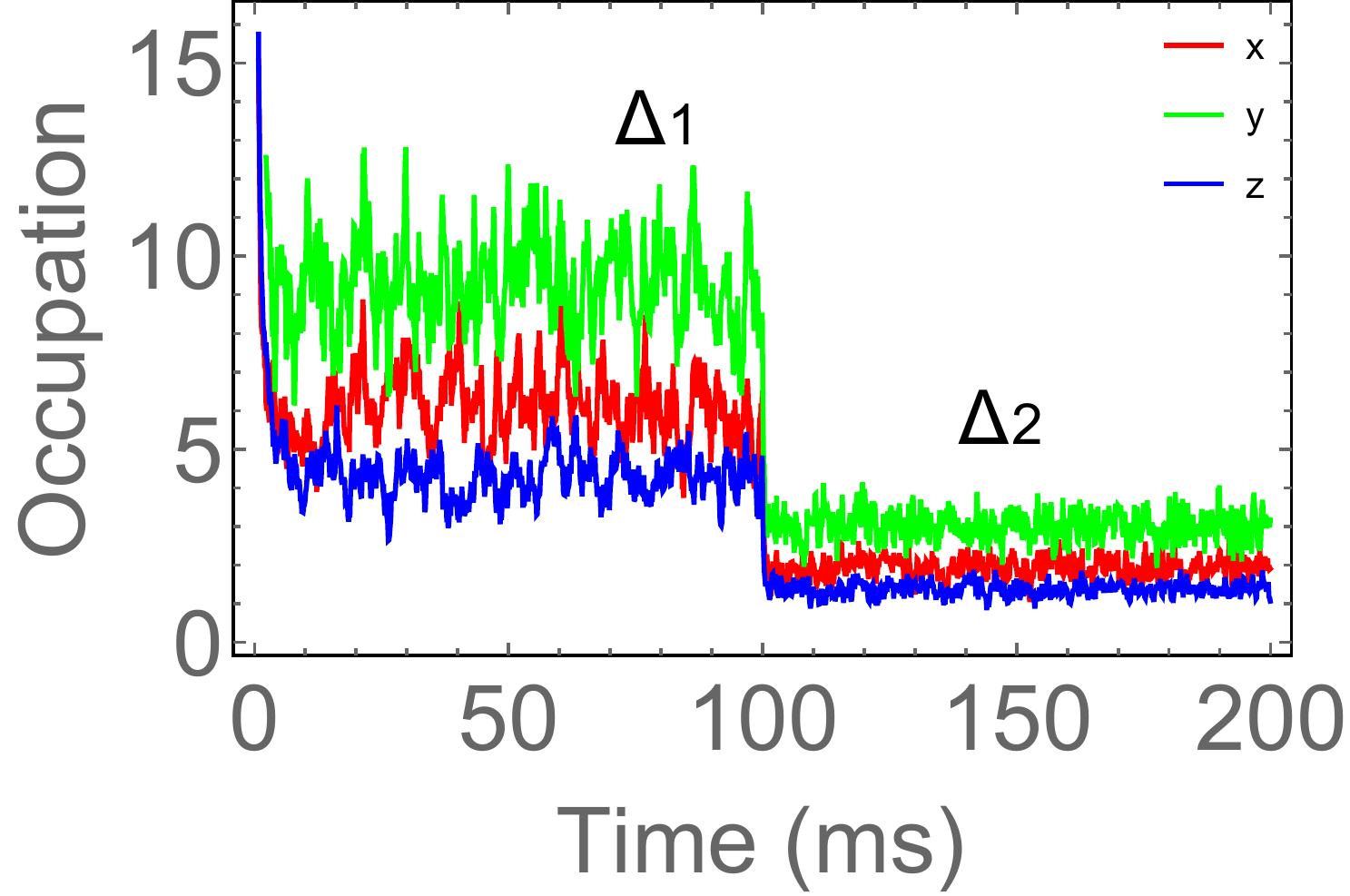}
\end{minipage}
\label{fg3a}
}
\subfigure
{
\put(22,34){(b)}
\begin{minipage}[]{0.22\textwidth}
\includegraphics[width=3.9cm,height=3.2cm]{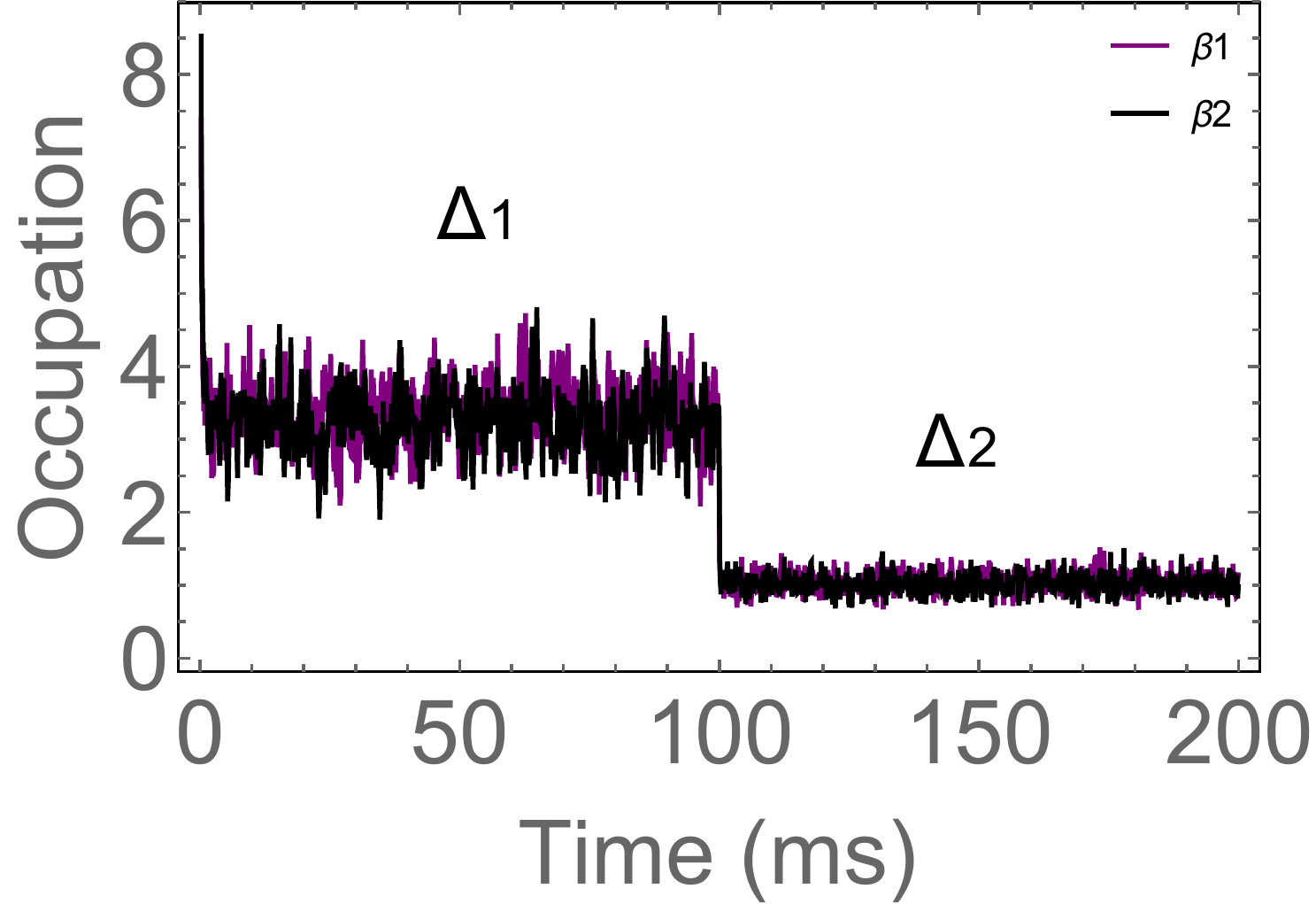}
\end{minipage}
\label{fg3b}
}
\subfigure
{
\put(29,34){(c)}
\begin{minipage}[]{0.24\textwidth}
\includegraphics[width=4.2cm,height=3.2cm]{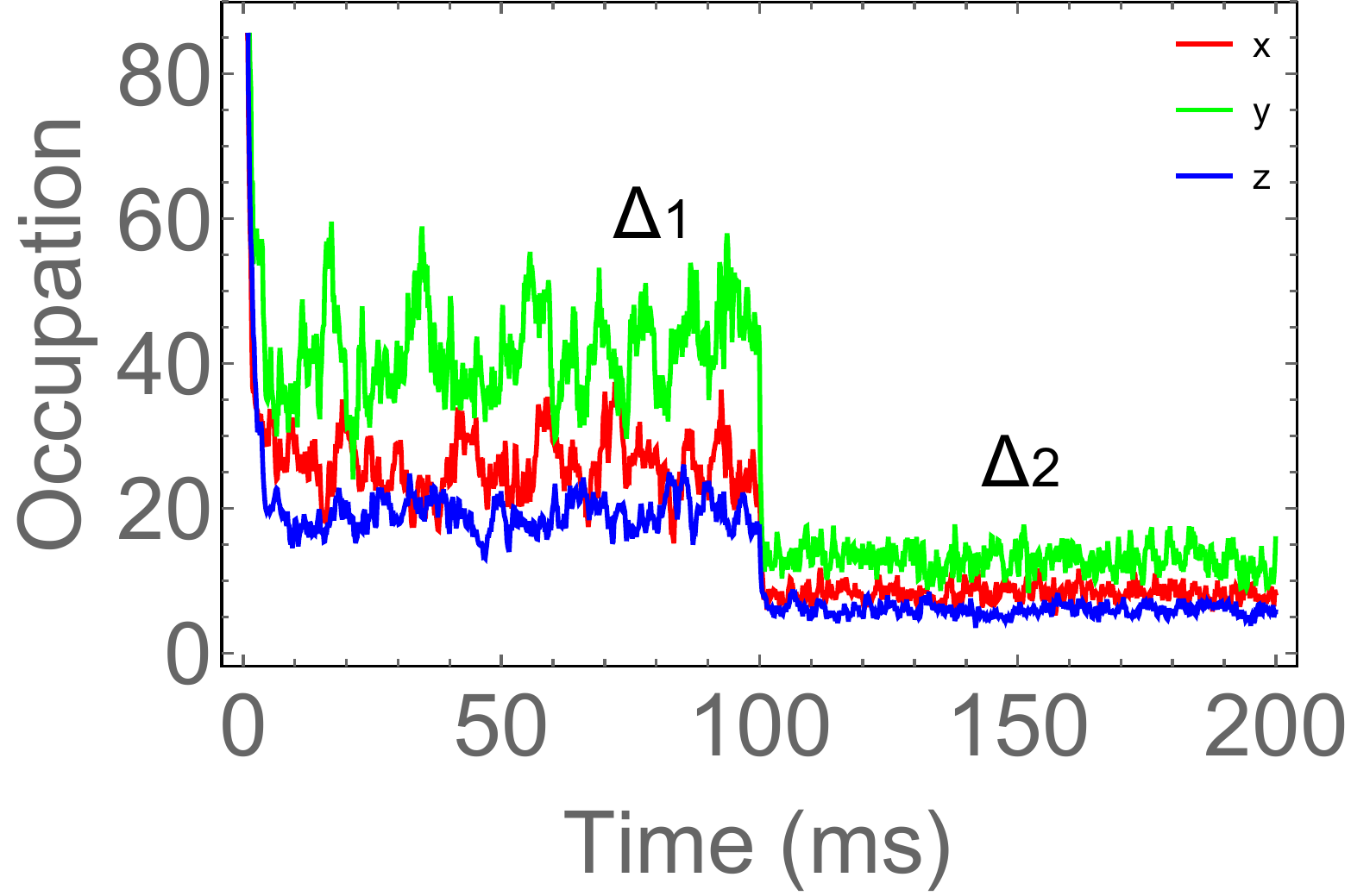}
\end{minipage}
\label{fg3c}
}
\subfigure
{
\put(24,34){(d)}
\begin{minipage}[]{0.22\textwidth}
\includegraphics[width=3.9cm,height=3.2cm]{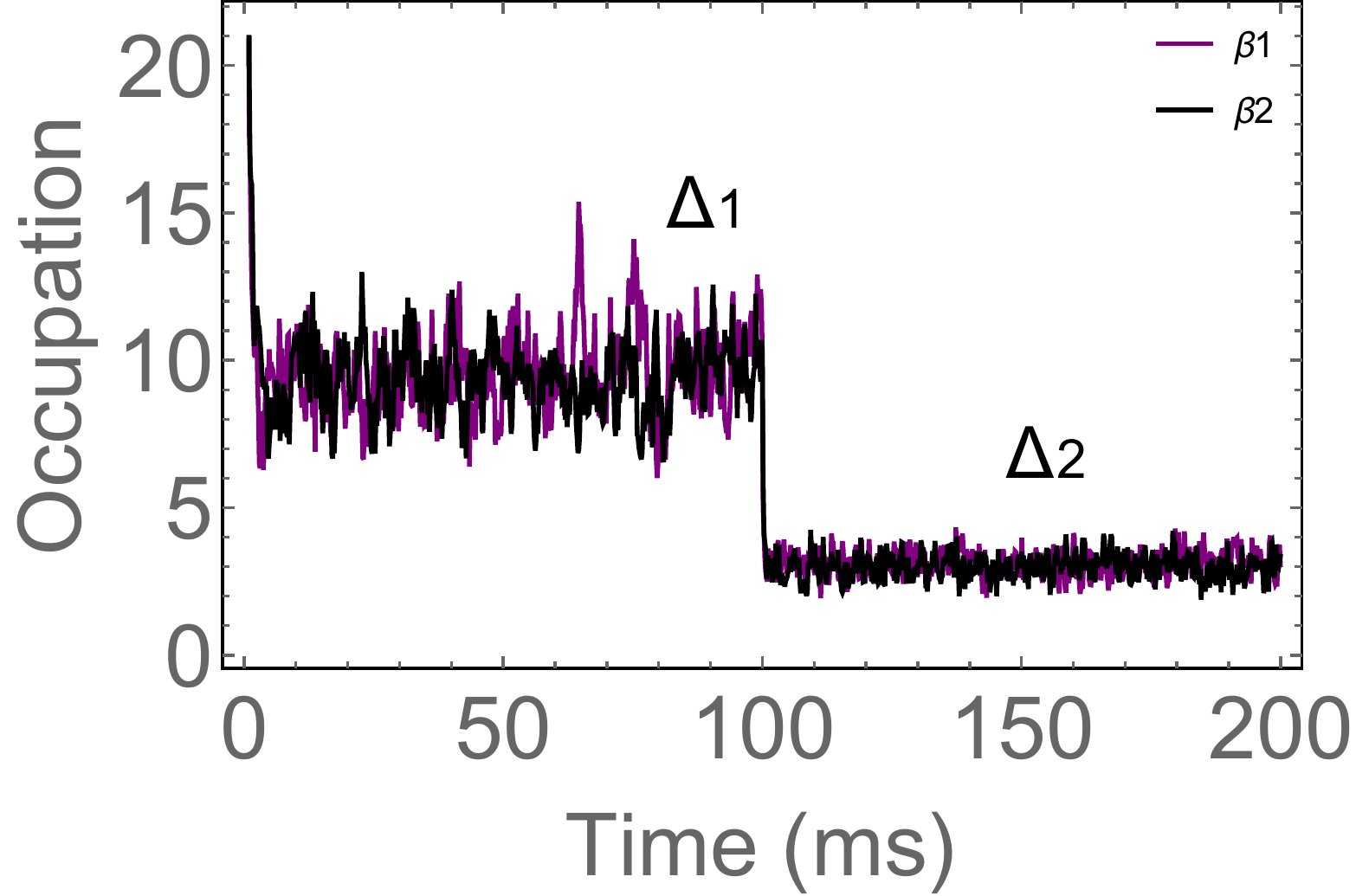}
\end{minipage}
\label{fg3d}
}
\subfigure
{
\put(29,34){(e)}
\begin{minipage}[]{0.24\textwidth}
\includegraphics[width=4.2cm,height=3.3cm]{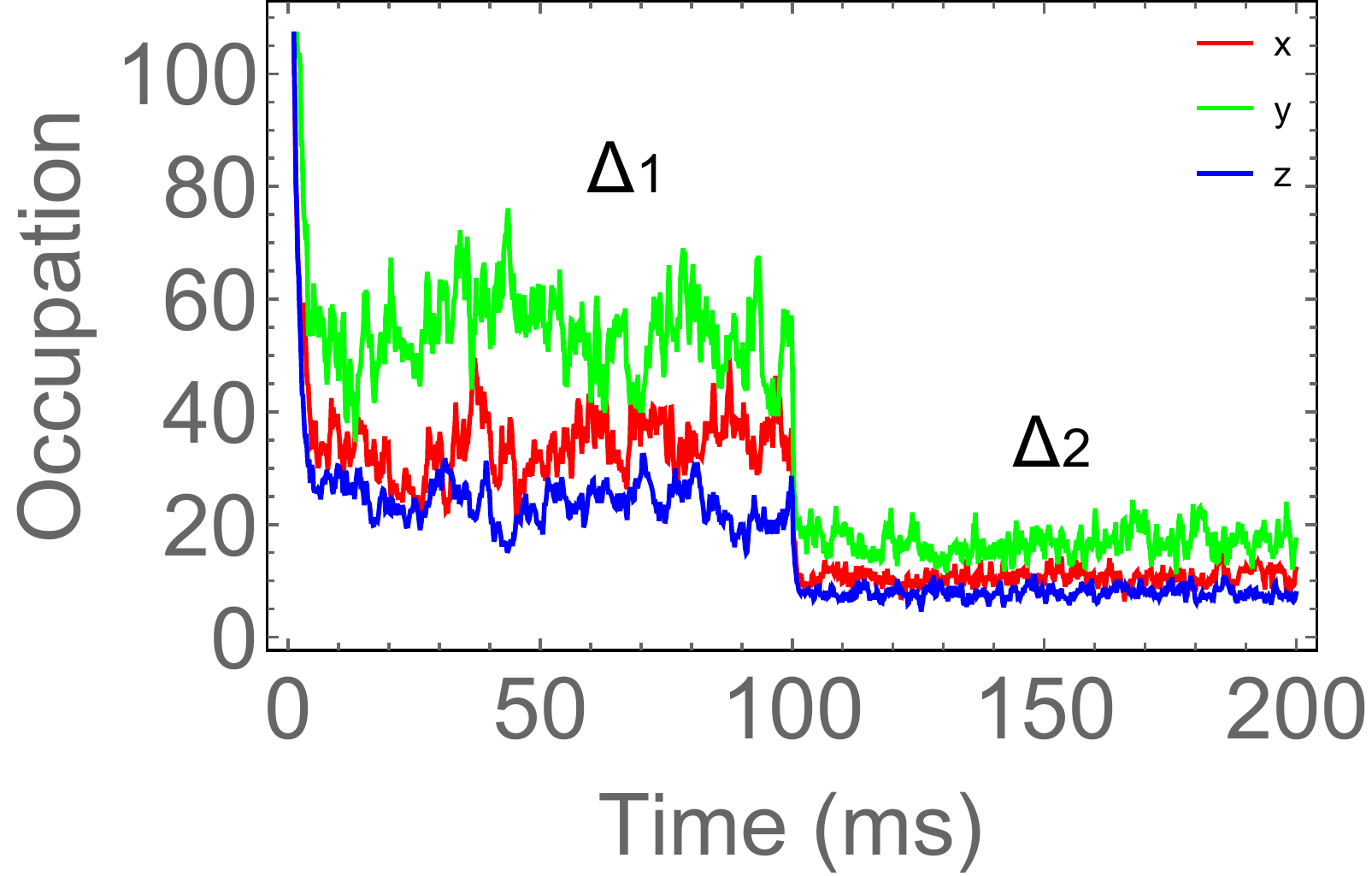}
\end{minipage}
\label{fg3e}
}
\subfigure
{
\put(28,34){(f)}
\begin{minipage}[]{0.22\textwidth}
\includegraphics[width=3.9cm,height=3.2cm]{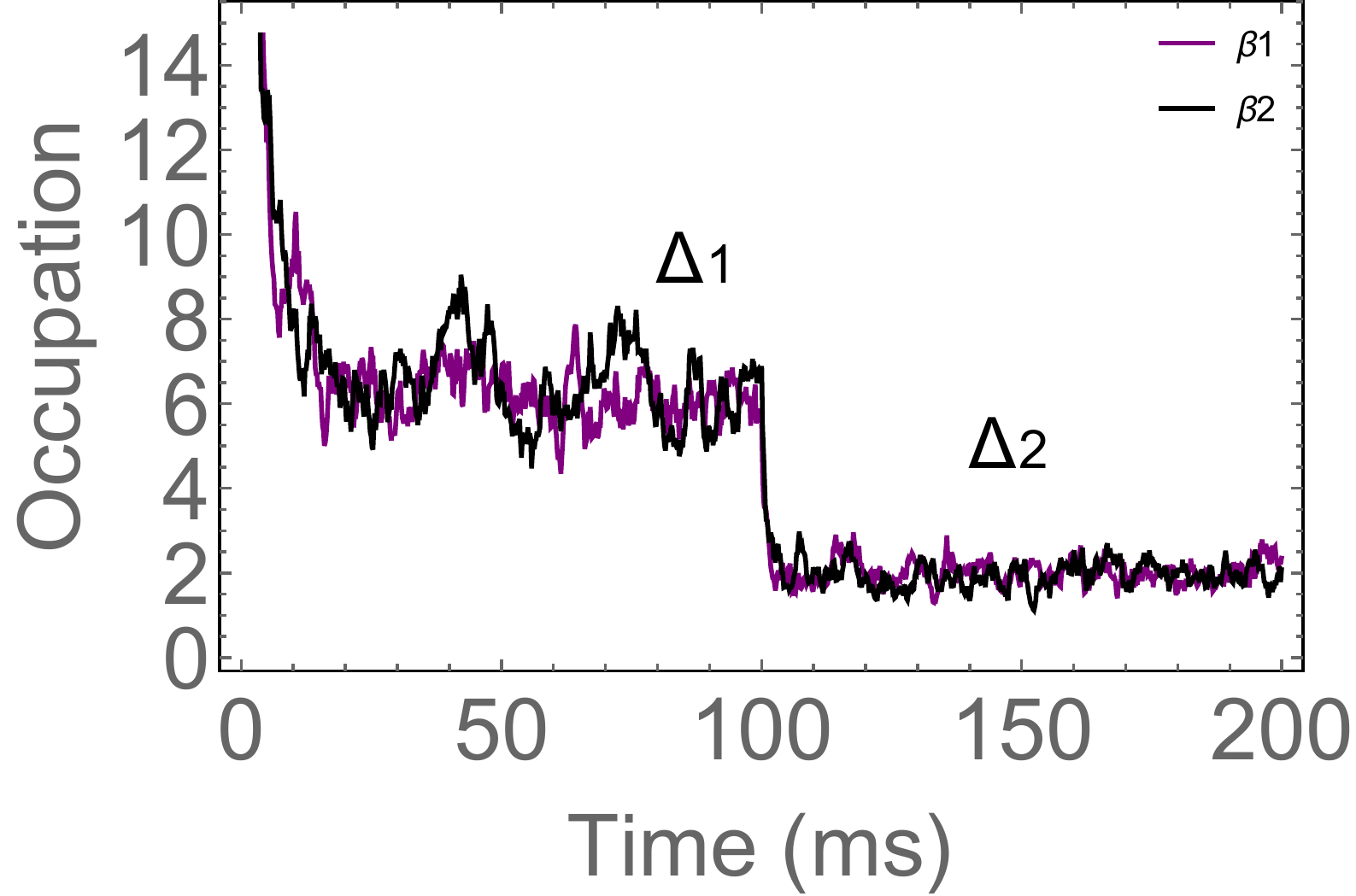}
\end{minipage}
\label{fg3f}
}
\caption{{(Color online)} The parametric feedback cooling for nano-diamonds in all degrees of freedom, where each curve shows the time evolution of the average occupation number in the corresponding degree of freedom. Data are collected by averaging $30$ cooling trajectories. Calculations are for classical parametric feedback cooling, thus results for occupation numbers less than 10 are suggestive. \textbf{(a)} and \textbf{(b)} depict the translational and rotational cooling respectively for a nanoparticle with half axes $(a=15\text{ nm},b=70\text{ nm})$. The cooling parameter $\Delta_1=\{\eta_i=1.1\times10^{11}\text{ }\text{s/m}^2,\zeta_i=10^{11}\text{ }\text{s/m}^2\}$ for $t<100$ms and $\Delta_2=10\Delta_1$ for $t>100$ms. Similarly, \textbf{(c)} and \textbf{(d)} show the cooling for half axes $(a=38\text{ nm},b=60\text{ nm})$ while \textbf{(e)} and \textbf{(f)} for half axes $(a=48\text{ nm},b=53\text{ nm})$. \label{fg3} }
\end{figure}

\begin{figure*}
\centering
\subfigure
{
\put(24,31){(a)}
\begin{minipage}[]{0.18\textwidth}
\includegraphics[width=3.2cm,height=2.8cm]{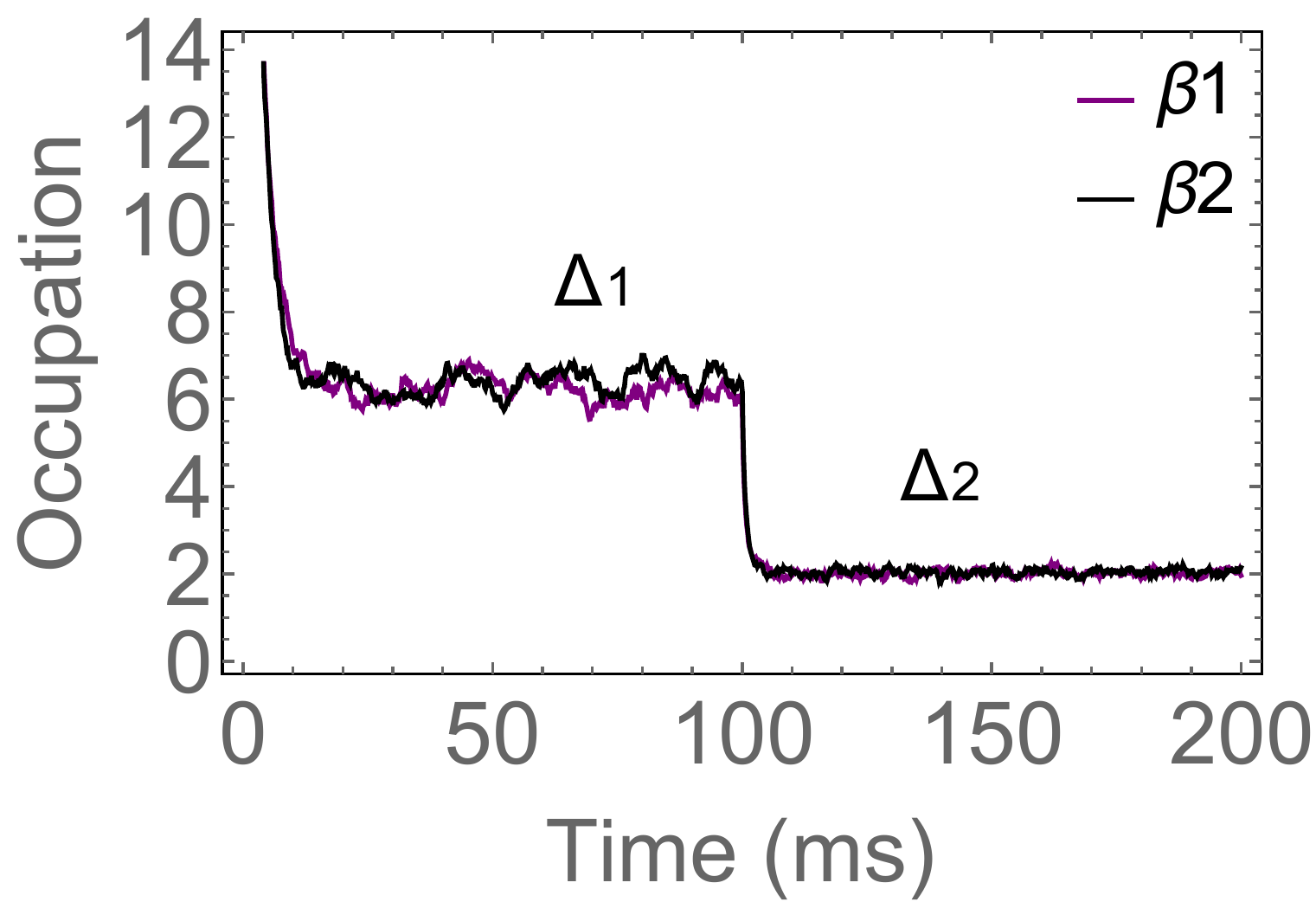}
\end{minipage}
\label{fg4a}
}
\subfigure
{
\put(24,31){(b)}
\begin{minipage}[]{0.18\textwidth}
\includegraphics[width=3.2cm,height=2.8cm]{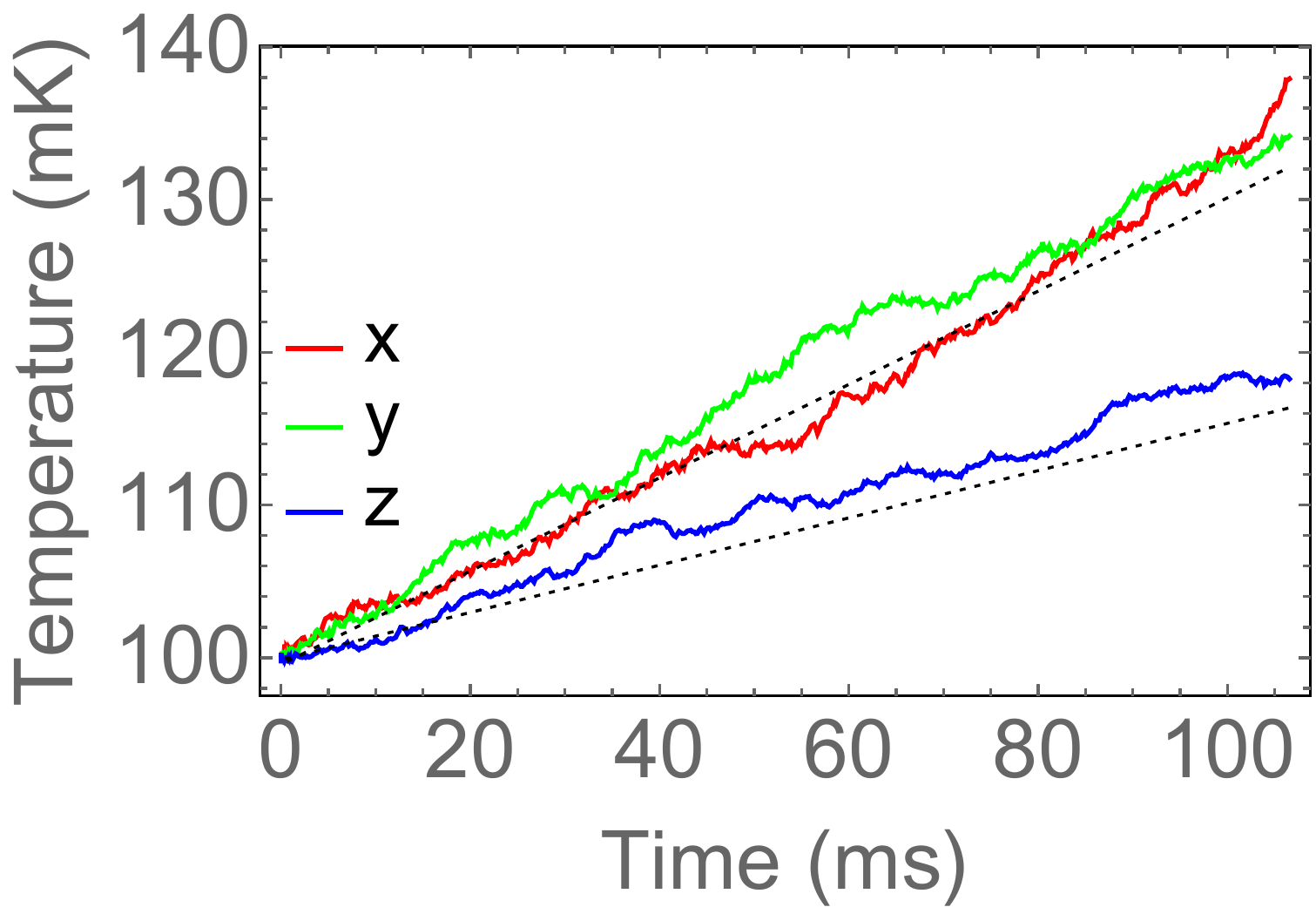}
\end{minipage}
\label{fg4b}
}
\subfigure
{
\put(24,31){(c)}
\begin{minipage}[]{0.18\textwidth}
\includegraphics[width=3.2cm,height=2.8cm]{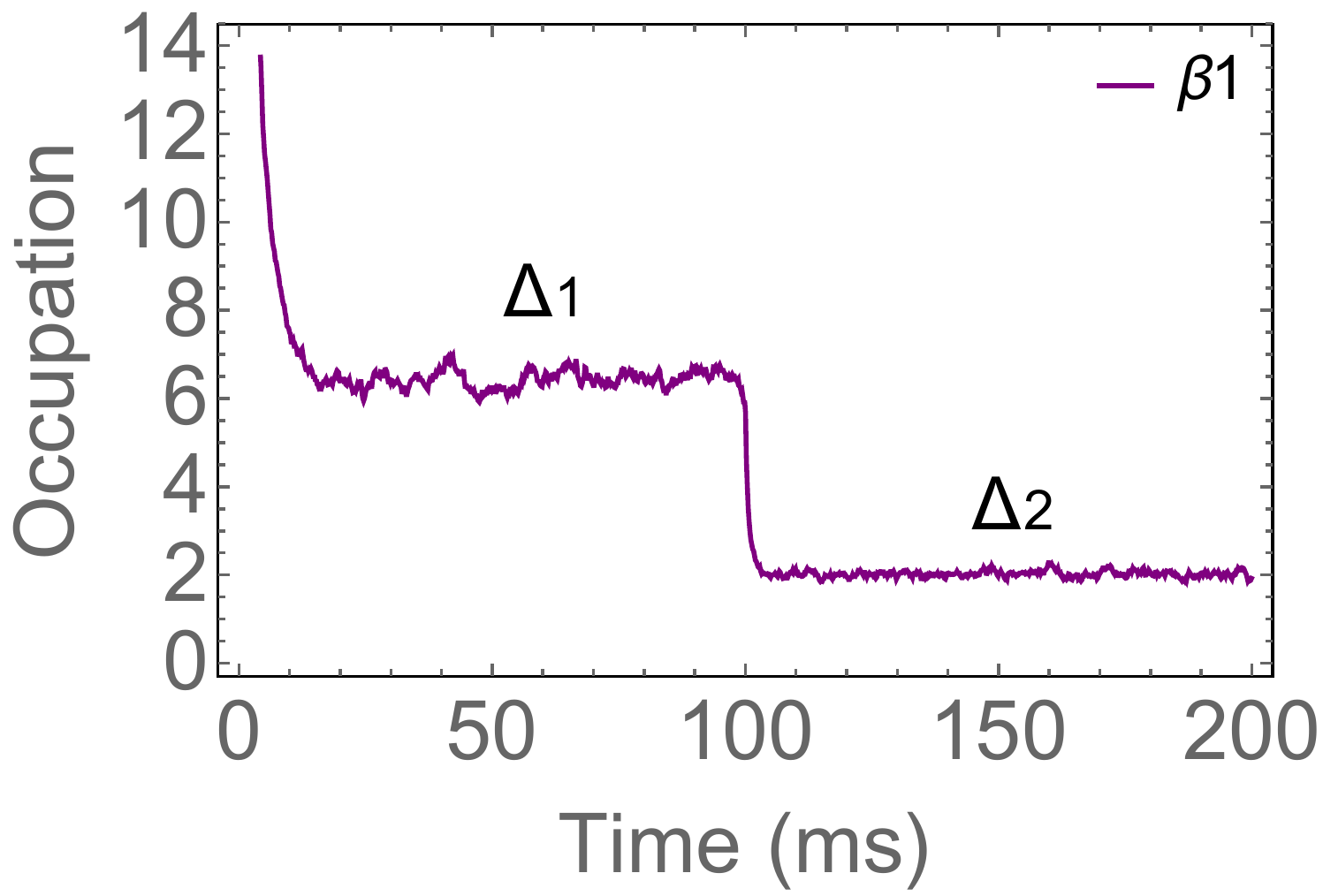}
\end{minipage}
\label{fg4c}
}
\subfigure
{
\put(24,28){(d)}
\begin{minipage}[]{0.18\textwidth}
\includegraphics[width=3.2cm,height=2.8cm]{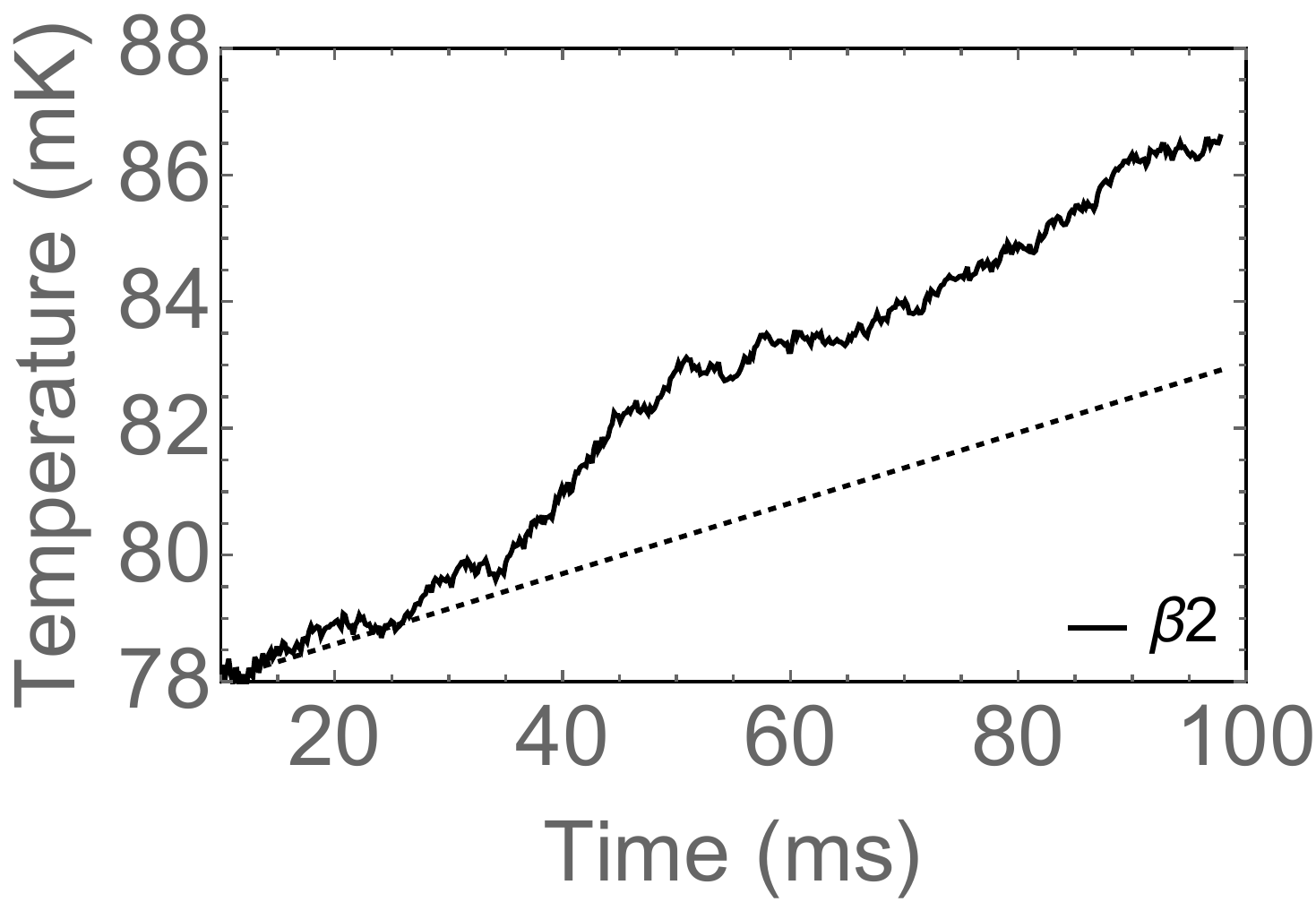}
\end{minipage}
\label{fg4d}
}
\subfigure
{
\put(24,28){(e)}
\begin{minipage}[]{0.18\textwidth}
\includegraphics[width=3.2cm,height=2.8cm]{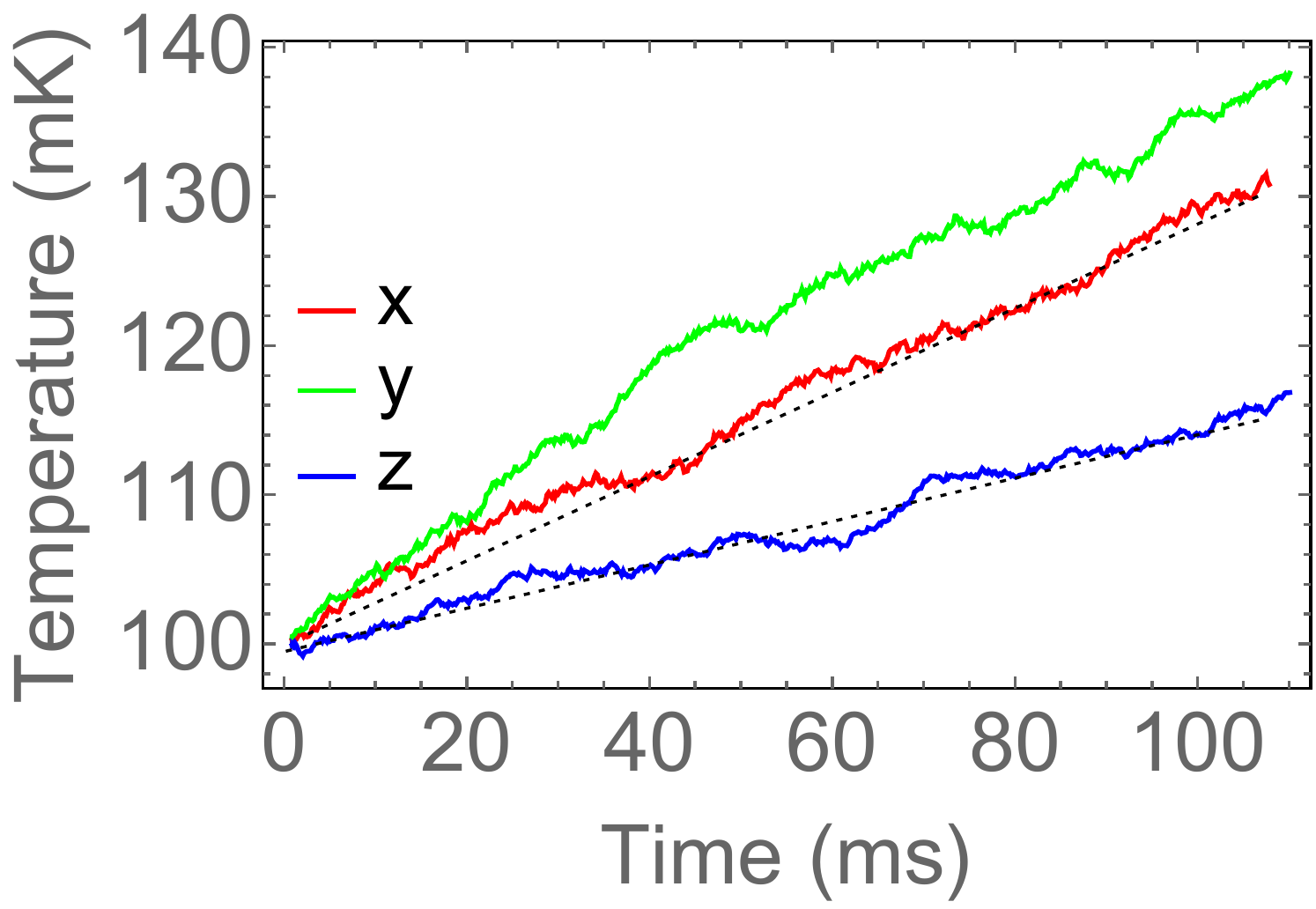}
\end{minipage}
\label{fg4e}
}

\caption{{(Color online)} The parametric feedback cooling in only the rotational degrees of freedom for a nano-diamond $(a=48\text{ nm},b=53\text{ nm})$. All curves are averaged over $400$ trajectories. \textbf{(a)} and \textbf{(b)} show the cooling in both $\beta_1$ and $\beta_2$ with the cooling parameters $\Delta_1=\{\eta_{1,2,3}=0,\zeta_i=10^{11}\text{ }\text{s/m}^2\}$. The black and purple lines show the rotational motion gets cooled as we increase the feedback parameters from $\Delta_1$ to $\Delta_2=10\Delta_1$. The red, green and blue lines depict that the heating trajectories in translational degrees of freedom. \textbf{(c)}, \textbf{(d)} and \textbf{(e)} show the result of cooling in only $\beta_1$ with parameters $\Delta_1=\{\eta_{1,2,3}=0,\zeta_2=0,\zeta_1=10^{11}\text{ }\text{s/m}^2\}$ and $\Delta_2=10\Delta_1$, and heating in $\beta_2$ and $x,y,z$ respectively. In \textbf{(d)}, resonance heating causes massive heating in the uncooled $\beta_2$ degree of freedom. The dashed lines in \textbf{(b)}, \textbf{(d)} and \textbf{(e)} are the heating curves $T=T_0+\dot{E} t$ with $T_0$ the initial temperature and $\dot{E}$ the corresponding heating rate from Tab. \ref{tab1}. \label{fg4}}
\end{figure*}

\section{Numerical simulation of shot noise heating and feedback cooling} \label{s4}

Parametric feedback cooling is discussed in Ref. \cite{SPFC}, where a single laser beam is used for both trapping and cooling. The spatial motion of a nanoparticle is cooled from room temperature to subkelvin, and the quantum ground state cooling is also suggested with the same cooling mechanism. In this parametric feedback scheme, a signal at twice the oscillation frequency is obtained by multiplying the particle's position with its first time derivative $x(t)\dot{x}(t)$. This information is then fed back to the system, which leads to a loop that on average acts as a drag on the particle. The parametric cooling works by simply modulating the intensity of the trapping laser, and this scheme is extremely suitable for rotational cooling since it avoids relatively complex operations if one tries to feedback torque. In this section, the feedback cooling calculations are based on ideal assumptions about measuring the nanoparticle's position and orientation. The discussions of feedback cooling with the measurement uncertainty are given in the next section.

Combining the translational and rotational motion, the classical dynamics of the ellipsoid is governed by
\begin{equation}
\begin{split}
\label{e28}
m\frac{d^2x_i}{dt^2}&\simeq-m\omega_i^2(1+\Delta)x_i,\\
I_1\frac{d^2\beta_j}{dt^2}&\simeq-I_1\omega_{\beta_j}^2(1+\Delta)\beta_j.
\end{split}
\end{equation}
The small oscillation approximation is used in the above equations where all corrections quadratic in the amplitude of oscillations have been dropped. $x_i=(x,y,z)$ and $\beta_j=(\beta_1,\beta_2)$. The shot noises in translation and rotation are added at each time step according to
\begin{equation}
\begin{split}
p_i(t+\delta t)&=p_i(t) +\delta W_i\cdot\delta p_i,\\
L_i(t+\delta t)&=L_i(t) +\delta R_j\cdot \delta L_j,
\end{split}
\end{equation}
where $\delta W_i,\delta R_j$ are the standard normally distributed random numbers, and $\delta p_i=\sqrt{2\dot{E}_{T_i}\delta t\cdot m}, \delta L_j=\sqrt{2\dot{E}_{R_j}\delta t\cdot I}$ are the fluctuation of the momentum and angular momentum for each degree of freedom induced by the shot noise. The heating rate in the $z$ direction (optical polarization direction) is half that of the other two translational degrees of freedom because the photons scatter less in the direction of the laser polarization \cite{DMOP}. $\Delta$ is a scalar which takes the form
\begin{equation}
\label{e30}
\Delta=\sum_{i=1,2,3}\eta_i x_i\dot{x}_i+\sum_{i=1,2}\zeta_i r^2\beta_i\dot{\beta}_i,
\end{equation}
where $r$ is the size of the nanoparticle. The feedback parameters $\eta_i$ and $\zeta_i$ have the unit $\text{Time/Length}^2$ and they control the cooling limit and speed. Details about the parameters and the parametric feedback cooling limit are given in the appendix. Simulations are performed for three different nano-diamonds with decreasing ellipticity, whose half axes go from ($a=15\text{ nm},b=70\text{ nm}$), ($a=38\text{ nm},60\text{ nm}$) to ($a=48\text{ nm},b=53\text{ nm}$). The corresponding parameters are given in Tab. \ref{tab1}. The classical equations of motion are numerically solved using a fourth-order Runga-Kutta algorithm with adaptive time steps \cite{NRIC}. All simulations are repeated many times and data is collected by averaging over the different runs to reduce the random noise.

We start by presenting the simulation with zero feedback ($\eta_{1,2,3}=0$, $\zeta_{1,2}=0$), which corresponds to the pure shot noise heating process. The system is prepared initially at temperature $T_i=1\text{ }\mu$K. The result is shown in Fig. (\ref{fg2}), where each curve depicts the time evolution of the energy in the corresponding degrees of freedom. Figure \ref{fg2a} and \ref{fg2b} show the case ($a=15\text{ nm},b=70\text{ nm}$) in the first $100$ ms. The rotational shot noise is about an order of magnitude larger than that in the translational motion. The case ($a=38\text{ nm},b=60\text{ nm}$) is given in Fig. \ref{fg2c} and \ref{fg2d}, in which the rotational and translational shot noise heating rates are of similar size. As the ellipticity gets smaller, the shot noise in the rotational degrees of freedom becomes less than that in the translational motion, which is shown in the Fig. \ref{fg2e} and \ref{fg2f} for the case ($a=48\text{ nm},b=53\text{ nm}$). From Tab. \ref{tab1}, the case ($a=48\text{ nm},b=53\text{ nm}$) has a higher rotational than translational oscillating frequency, which suggests that it might be a good candidate for rotational cooling. 

The non-zero feedback cooling is performed with the system temperature initially prepared at $T_i=0.1\text{ K}$. The feedback parameters ($\eta_i$, $\zeta_i$) are chosen in a way such that Eq. (\ref{e30}) is much smaller than one and the position and velocity are assumed to be measured perfectly. 

First, we turn on the feedback in all degrees of freedom. The results are shown in Fig. (\ref{fg3}). Because the calculations are classical, the results for occupation less than ~10 are qualitative/suggestive. However, we do expect that the classical results are approximately correct for $\braket{n}\sim10$ so we do expect this feedback could get to near the ground state. By tuning the feedback parameters from $\Delta_1=\{\eta_{1,2,3}=1.1\times10^{11}\text{ }\text{s/m}^2, \zeta_{1,2}=10^{11}\text{ }\text{s/m}^2\}$ to $\Delta_2=10\Delta_1$, the system is observed to be quickly cooled. Both the translational and rotational occupation numbers can get down to less than one in this classical calculation, which suggests a possibility of ground state cooling in all degrees of freedom. Figure \ref{fg3a} and \ref{fg3b} depict the cooling of a nano-diamond with half axes $(a=15\text{ nm},b=70\text{ nm})$ in the translational and rotational degrees of freedom respectively. We see that rotation and translation are cooled with almost equal speed though the rotational oscillating frequency is more than six times higher than that for the translational motion. As the ellipticity goes lower, the cooling in rotation becomes more effective than in translation. As shown in Fig. \ref{fg3c} and \ref{fg3d} for case $(a=38\text{ nm},b=60\text{ nm})$, when the parameter $\Delta_1$ is taken, the rotational occupation numbers go down close to 10 while the translational occupation numbers are still around 20. The cooling in rotation gets even better when the particle with half axes $(a=48\text{ nm},b=53\text{ nm})$ is used, where the rotation is close to the ground state ($\braket{n}<1$) while the translational occupation numbers are still more than 10, as shown in Fig. \ref{fg3e} and \ref{fg3f}. The reason is that when the ellipticity of the nanoparticle gets smaller, the rotational shot noise heating is less than that for translational heating while the rotational oscillating frequency is still larger than that for translation. Thus, a better rotational cooling for a particle with low ellipticity is expected, which was suggested in the previous section. From the appendix, the steady state value of $\braket{n}$ is proportional to the square root of $\dot{E}/\omega^2$. This suggests that smaller values of $\Delta n=2\pi\dot{E}/(\hbar\omega^2)$, as defined in the previous section, are better for cooling to the ground state. However, we will see in the next section that measurement noise qualitatively modifies this trend.

Second, we keep the feedback cooling only in the rotational degrees of freedom with $\Delta_1=\{\eta_{1,2,3}=0, \zeta_{1,2}=10^{11}\text{ }\text{s/m}^2\}$ and $\Delta_2=10\Delta_1$. The results are shown in Fig. \ref{fg4a} and \ref{fg4b} for the nano-diamond with half axes $(a=48\text{ nm},b=53\text{ nm})$. As shown in Fig. \ref{fg4a}, when the feedback is increased from $\Delta_1$ to $\Delta_2=10\Delta_1$, the rotational occupation number goes down all the way to the quantum regime. However, as shown in Fig. \ref{fg4b}, the translational motion is heated up in the mean time. In order to see the cooling in only one rotational degree of freedom, we also calculate the case with $\Delta_1=\{\eta_{1,2,3}=0, \zeta_2=0,\zeta_1=10^{11}\text{ }\text{s/m}^2\}$. As shown in Fig. \ref{fg4c}, \ref{fg4d} and \ref{fg4e}, the motion in $\beta_1$ degree of freedom quickly gets cooled to the ground state regime when $\Delta_2$ are taken, while all other degrees of freedom ($\beta_2$, $x,y$ and $z$) are heated up. For $\beta_2$, extra heating is observed due to the resonance heating: the changes in the laser intensity are predominantly at the frequency to resonantly couple with either of the rotational degrees of freedom. In Fig. \ref{fg4b}, \ref{fg4d} and \ref{fg4e}, the dashed lines show the heating from pure shot noise. We see that the pure shot noise heating rates are slightly lower (almost the same) than the heating rate with feedback cooling. The reason is because the cooling in one degree of freedom can add to the heating in the other degrees of freedom. Fortunately, this extra heating is not excessive and should not be a problem in experiments.

\section{The parametric feedback cooling limit with classical uncertainty} \label{s5}

The above discussion of feedback cooling is based on an ideal measurement of the particle's position and velocity. In reality, a measurement can't be infinitely accurate and is fundamentally limited by the quantum uncertainty $\delta x\delta p\geqslant\hbar/2$, which introduces an extra feedback noise during the cooling process. The uncertainty in the position measurement can be reduced by increasing the photon scattering rate, however stronger photon scattering induces faster shot noise heating. Thus, tuning an appropriate photon recoil rate and a proper feedback parameter should be important in optimizing the feedback cooling.

This section numerically studies the optimal cooling limit when the main error in the position measurement is due to classical measurement uncertainty. As we will show below, the equations of motion can be scaled. Therefore, the simulation is performed in only the $x$ degree of freedom for the case ($a=48\text{ nm},b=53\text{ nm}$) in Tab. \ref{tab1}. The calculation is still classical, but the feedback signal is modified to satisfy $\delta x\delta p = N\hbar/2$, where $N$ is a measure of the classical uncertainty. The dynamical equation is given by 
\begin{equation}
\begin{split}
m\frac{d^2x}{dt^2}&=-m\omega_x^2(1+\eta x_m\dot{x}_m)x,\\
x_m&=x+\delta R\cdot\delta x,
\end{split}
\end{equation}
and the shot noise is added according to
\begin{equation}
p(t+\delta t)=p(t)+\delta W\cdot\delta p,
\end{equation}
where $\delta R$ and $\delta W$ are Gaussian random numbers with unit variance, $\delta p=\sqrt{2\dot{E}_{T_x}\delta t\cdot m}$ is the momentum fluctuation determined by the shot noise $\dot{E}_{T_x}$, and $x_m$ is the measured position with $\delta x=N\hbar/(2\sqrt{2\dot{E}_{T_x}\delta t\cdot m})$, which is chosen to satisfy the relation $\delta x\delta p = N\hbar/2$. Several values of $N$ are used in the the pure classical calculation. In reality, the results are not physically possible for $N < 1$, and for small $N$ the result is only suggestive because it would require a true quantum treatment. Figure (\ref{fg6}) shows the results, where each curve corresponds to the steady state occupation in terms of the feedback parameter $\eta$. The pink line corresponds to the classical feedback with no noise in the position measurement ($N=0$), where the occupation keeps decreasing as we increase the feedback parameter. As we add uncertainty to the feedback signal, the purple ($N=1$), black ($N=1.5$), green ($N=2$), red ($N=2.5$) and blue ($N=3$) lines will go up after passing their minimal occupations, which are the corresponding optimal cooling limit. The reason is that, as $\eta$ increases, the feedback cooling is strengthened, but the noise in the measured value of $x$ leads to the feedback procedure itself adding noise to the motion. Beyond a value of $\eta$, the feedback noise heating becomes faster than the feedback cooling, which indicates that the steady state occupation can reach a minimum and then increase. Moreover, one can see that a larger uncertainty in the position measurement leads to a larger occupation for optimal cooling limit. The reason is that the feedback noise heating is generally faster with a big $N$ in the position measurement than that with a smaller $N$.

\begin{figure}
\includegraphics[width=7.0cm,height=5.0cm]{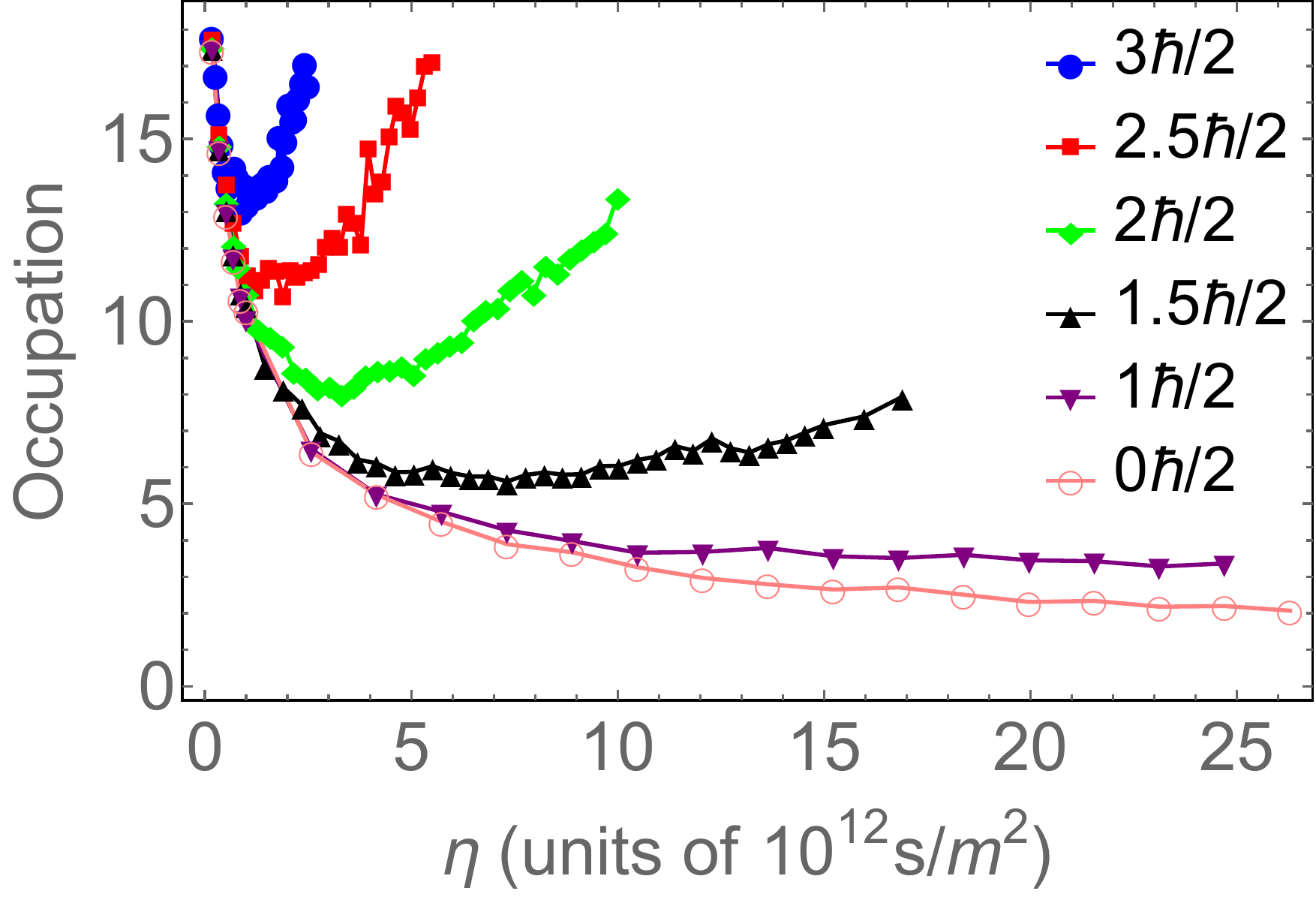}
\caption{ (Color online) The steady state occupation in terms of the feedback paramter $\eta$ for $x$ degree of freedom of the particle ($a=48\text{ nm},b=53\text{ nm}$) from Tab. \ref{tab1}. The different curves correspond to several different values of classical uncertainty.  \label{fg6}}
\end{figure}

The steady state occupation is also related to the shot noise heating and the oscillating frequency, as suggested by the result in the appendix $\braket{n}_{limit}\propto\sqrt{\dot{E}_T/\omega^2}$ for ideal measurements. In fact, the one dimensional dynamical equation for the nanoparticle can be scaled 
\begin{equation}
\begin{split}
\label{e36}
\frac{d^2\tilde{x}}{d\tilde{t}^2}&=-\tilde{x}(1+\eta\frac{\hbar}{2m}\tilde{x}_m\dot{\tilde{x}}_m),\\
\tilde{x}_m&=\tilde{x}+\delta R\cdot \delta\tilde{x},
\end{split}
\end{equation}
and the shot noise is added according to
\begin{equation}
\label{e37}
\tilde{p}(t+\delta t)=\tilde{p}(t)+\delta\tilde{p}(t)\delta W,
\end{equation}
where the scaled position $\tilde{x}=x/a_0$ with $a_0=\sqrt{\hbar/(2m\omega_x)}$, $\tilde{t}=\omega_xt$, and $\dot{\tilde{E}}_T=2\dot{E}_T/(\hbar\omega_x^2)$. $\delta\tilde{p}=\sqrt{{2\dot{\tilde{E}}_T}\cdot{d\tilde{t}}}$ and $\delta\tilde{x}=N\sqrt{1/(2\dot{\tilde{E}}_Td\tilde{t})}$. The scaled equation shows that $\Delta n=2\pi\dot{E}_T/(\hbar\omega_x^2)$ (as defined previously), $N$ and $\eta$ determine the particle's dynamics. To confirm that, we simulate the cooling of the $x$ degree of freedom for particle ($a=48\text{ nm},b=53\text{ nm}$) with fixed measurement uncertainty ($N=2$). First, we choose ($\dot{E}_T$, $\omega_x$) to be different values (470 mK/s,343 kHz), (824 mK/s,454 kHz) and (1295 mK/s,569 kHz), which are obtained by tuning the laser power to $P=(40\text{ mW},70\text{ mW},110\text{ mW})$ respectively. Figure \ref{fg7a} gives the simulation results, where the three curves give the steady state occupation in terms of the feedback parameter. Those curves match each other, which confirms that $\Delta n$ indeed determines the dynamics, since varying the laser power doesn't change the quantity $\Delta n\simeq0.083$. All three curves get to an optimal cooling limit around $\braket{n}=8.5$ when $\eta=3.3\times10^{12}\text{ }\text{s/m}^2$. In Fig. \ref{fg7b}, we take $\Delta n\simeq0.026$ by changing the laser beam waist. Using the same laser powers $P=(40\text{ mW},70\text{ mW},110\text{ mW})$, the shot noise heating and the $x$ translational oscillating frequency are (1488 mK/s,1085 kHz), (2603 mK/s,1434 kHz) and (4092 mK/s,1798 kHz) respectively. The three curves still match, but the minimal point is shifted to ($\braket{n}=12,\eta=5\times10^{11}\text{ }\text{s/m}^2$), which suggests that the optimal cooling limit should depend on the choice of $\Delta n$ for a given value of $N$, the scale factor between the uncertainty in the position measurement, $\delta x$, and the momentum shot noise scale, $\delta p$. 

Comparing the two results in Fig. (\ref{fg7}), we see that a lower optimal cooling limit is reached for the motion with a bigger $\Delta n$ when $N$ is held fixed. This motivates us to calculate the optimal cooling limit for varied $\Delta n$ (by changing the beam waist), and the result is show in Fig. (\ref{fg8}), where the two curves correspond to $N=(1,2)$. Both curves reveal that a bigger $\Delta n$ leads to a lower optimal cooling limit, which suggests that a more accurate feedback cooling can beat the cost from the higher shot noise heating for fixed $N$. The fact that a higher shot noise leads to a lower optimal occupation might be because of (1) a higher shot noise indicates a more accurate and effective feedback cooling; (2) an accurate feedback induces a lower feedback noise. Figure (\ref{fg8}) also shows that a smaller $N$ generally has a lower optimal cooling limit, which matches the result in Fig. (\ref{fg6}). The data in Fig. (\ref{fg8}) stops at $\Delta n=0.41$, since a bigger $\eta$ is needed in order to get to the optimal cooling limit. Our calculation becomes unstable when $\eta$ gets larger. In reality, a bigger feedback parameter $\eta$ means a lot more effort in feedback cooling. The maximal realizable $\eta$ in the experiment should physically bound the lowest cooling limit for fixed $N$. The actual shot noise heating rate and measurement uncertainty determines the minimum occupation number. By scaling these parameters, one can understand how the system will respond in terms of the dimensionless Eqs. (\ref{e36}) and (\ref{e37}).

\begin{figure}
\centering
\subfigure
{
\put(29,35){(a)}
\begin{minipage}[]{0.23\textwidth}
\includegraphics[width=3.9cm,height=3.2cm]{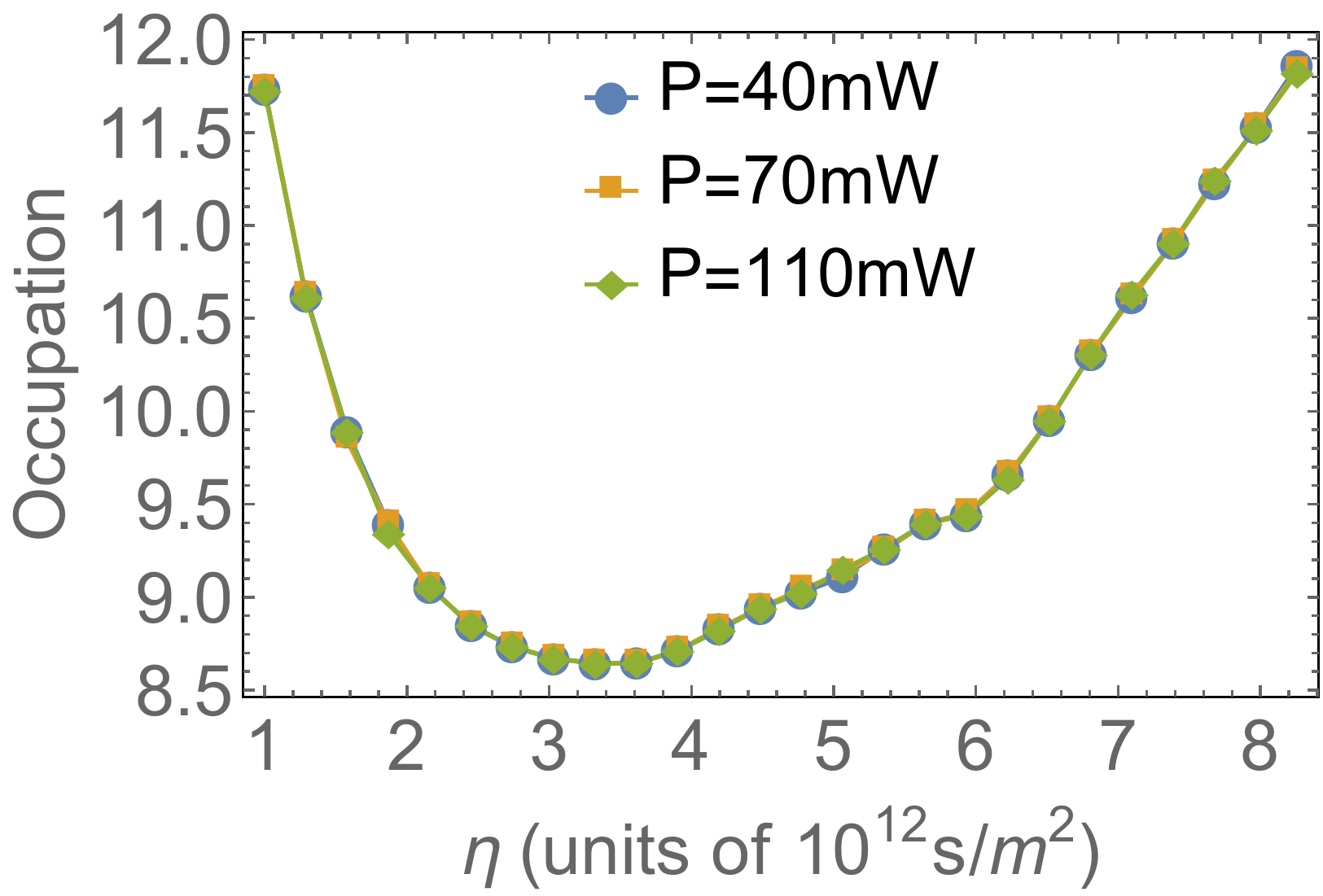}
\end{minipage}
\label{fg7a}
}
\subfigure
{
\put(24,35){(b)}
\begin{minipage}[]{0.23\textwidth}
\includegraphics[width=3.9cm,height=3.2cm]{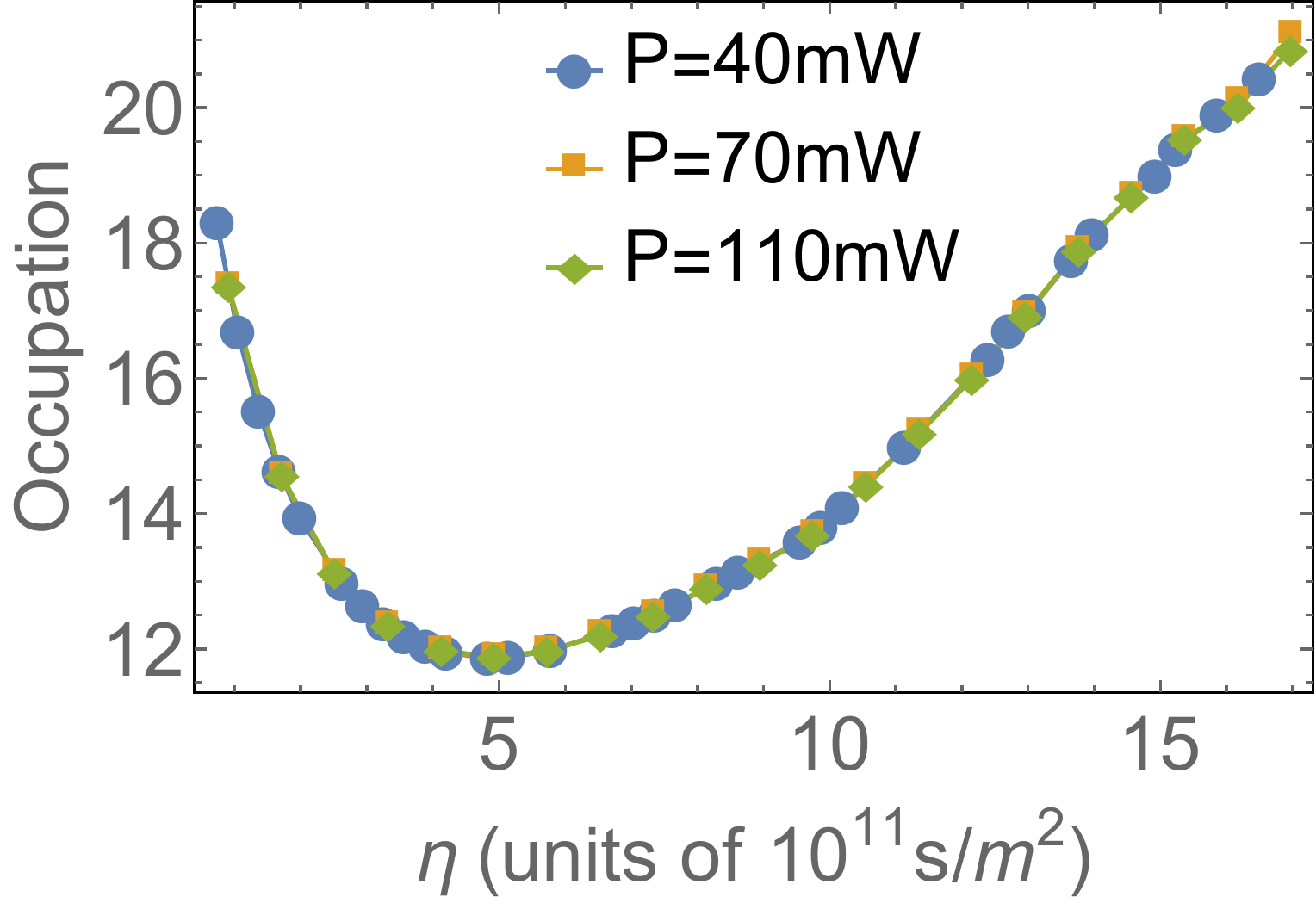}
\end{minipage}
\label{fg7b}
}
\caption{{(Color online)} The steady state occupation for $x$ degree of freedom of particle ($a=48\text{ nm},b=53\text{ nm}$) from Tab. \ref{tab1} in terms of the feedback parameter $\eta$ with $N=2$. Three different laser powers $P=(40\text{ mW},70\text{ mW},110\text{ mW})$ are used. \textbf{(a)} The quantity $\Delta n=0.083$; \textbf{(b)} The quantity $\Delta n=0.026$.
\label{fg7} }
\end{figure}

\begin{figure}
\includegraphics[width=7.0cm,height=5.0cm]{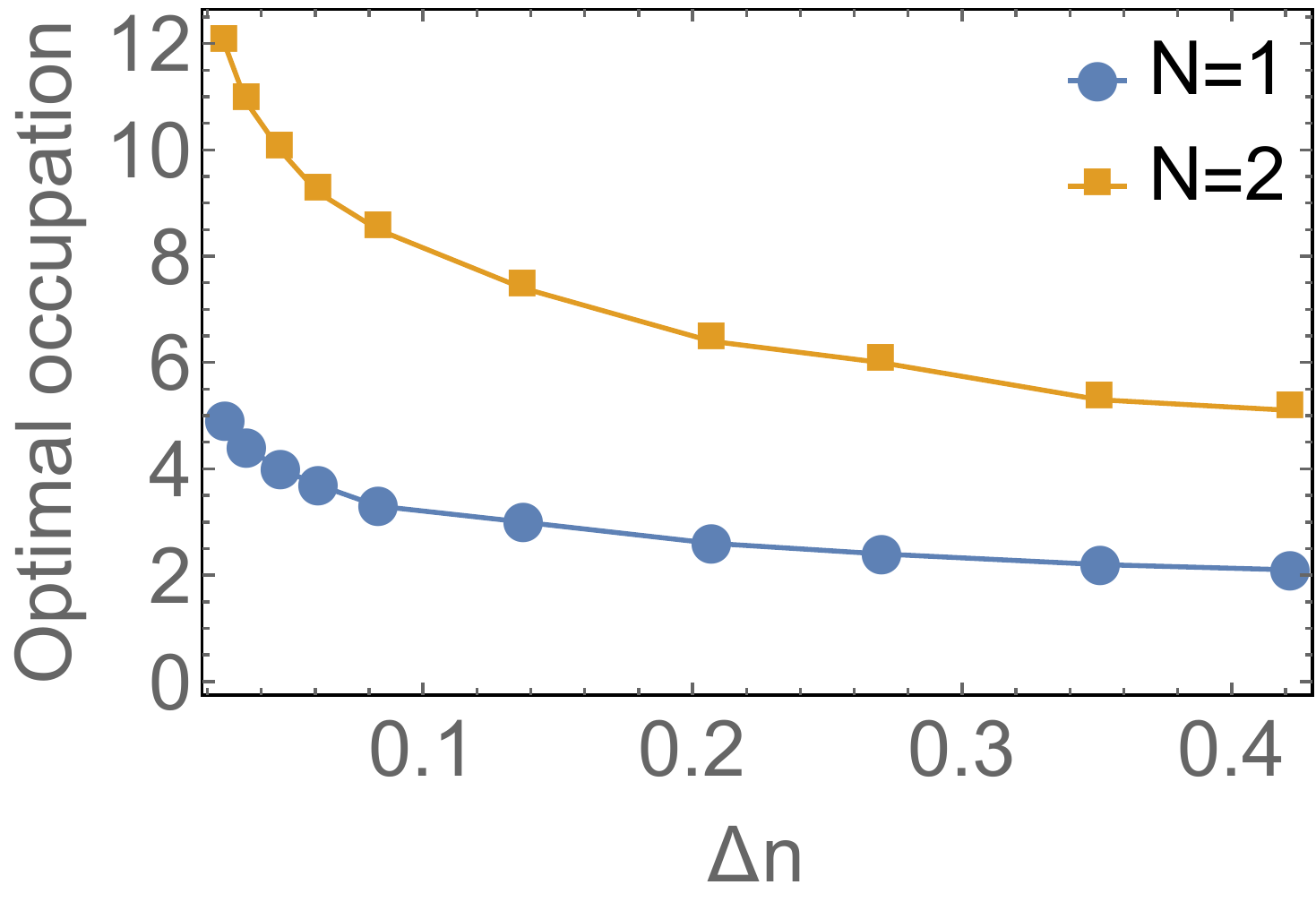}
\caption{ (Color online) The optimal cooling limit for $x$ degree of freedom of particle ($a=48\text{ nm},b=53\text{ nm}$) from Tab. \ref{tab1} with respect to $\Delta n$. The blue and yellow curves correspond to the classical uncertainty measure $N=1$ and $N=2$ respectively. Our data stops at $\Delta n=0.41$ since the feedback calculation with a larger $\eta$ becomes unstable when we try to reach the optimal cooling limit. \label{fg8}}
\end{figure}

\section{conclusion} \label{s6}

The translational and rotational shot noise heating and feedback cooling of an optically trapped nano-ellipsoid were analytically and numerically investigated. The detailed analysis suggests that a lower relative rotational heating rate is expected for a wide range of nanoparticle geometries. This conclusion is in contrast to that when scattering from black body radiation was studied \cite{DORD} which reported that rotational degrees of freedom decohered much faster than translational degrees of freedom. The qualitatively different conclusion is due to the difference in photon scattering from a polarized beam aligned along the nanoparticle axis compared to unpolarized photons. 

The analysis and numerical calculation of the shot noise heating suggest that a lower relative rotational heating rate results from (1), a nanoparticle with near to spherical shape for fixed size; (2), a nanoparticle with a bigger size for fixed ellipticity; (3), a trapping laser with a shorter wavelength and a bigger beam waist; (4), a nanoparticle with lower dielectric constant. In addition, the calculation of the feedback cooling in only the rotational degrees of freedom reveals that a separate rotational cooling should be experimentally possible, since heating in the other degrees of freedom was only slightly faster than the shot noise. 

The feedback cooling with classical measurement uncertainty was analyzed. The measurement uncertainty introduces an extra noise during the feedback, which competes with the cooling when the feedback parameter increases. When the scaled classical uncertainty $N$ is held fixed, a system with a bigger value of $\Delta n=2\pi\dot{E}/(\hbar\omega^2)$ could in principle get to a lower optimal cooling limit. While this is an interesting result, it is hard to imagine an experiment where the $N$ can be held fixed while the shot noise heating rate is changed as it would require the uncertainty in  $x$ to decrease proportional to $1/\sqrt{\dot{E}}$ as the heating rate increases. A more effective way to achieve small occupation number is to decrease $N$ which is proportional to the uncertainty in $x$ times $\sqrt{\dot{E}}$. 

In conclusion, the shot noise heating, the measurement uncertainty, and the feedback parameter are important factors to consider when cooling a levitated nanoparticle in the shot noise dominant region. The results presented here can provide a framework for thinking about how these parameters affect the heating and the feedback cooling of levitated nanoparticles. However, since our calculations are classical, there is clearly a need for investigations of quantum effects on feedback cooling for small occupation number. The results in Fig. (\ref{fg8}) suggest there may be non-intuitive trends in the quantum limit.

This work was supported by the National Science Foundation under Grant No.1404419-PHY.

\begin{appendix}

\section{The parametric feedback cooling}

This appendix describes the parametric feedback cooling scheme and analyzes the cooling limit in the shot noise dominant regime. Perfect measurement is assumed in the following derivation. As an example, the average cooling power for one translational degree of freedom from the feedback is given by
\begin{equation}
\braket{P}=-\eta k\braket{x^2\dot{x}^2}\simeq-\frac{\eta E^2}{2 m} .
\end{equation}
where $E$ is the system energy in this degree of freedom and $k$ is the spring constant. The approximation is made above by ignoring the noise when taking the cycle average. The negative sign of the power guarantees an effective cooling during the feedback process. Combining with the translational shot noise heating rate, the system energy follows the differential equation
\begin{equation}
\frac{dE}{dt}=\dot{E}_T-\frac{\eta E^2}{2 m}.
\end{equation} 
A steady state can be reached when the heating and cooling are balanced, which yields the cooling limit
\begin{equation}
\braket{n}_{limit}=\sqrt{\frac{2m\dot{E}_T}{\eta\hbar^2\omega^2}},
\end{equation}
where $\omega$ is the oscillation frequency. One finds that a bigger $\eta$ gives a lower steady state energy and the particle mass together with the quantity $\dot{E}_T/\omega^2$ determine the final occupation. The differential equation can be analytically solved 
\begin{equation}
E=E_{limit}\left(1+\frac{2}{B\exp(2\sqrt{\frac{\eta\dot{E}_T}{2m}}t)-1}\right),
\end{equation}
where 
\begin{equation}
B=\frac{\sqrt{\eta/2m}E_i+\sqrt{\dot{E}_T}}{\sqrt{\eta/2m}E_i-\sqrt{\dot{E}_T}}.
\end{equation}
$E_i$ is the initial energy of the system. The system gets cooled as time increases and the parameter $\sqrt{\frac{\eta\dot{E}_T}{2m}}$ is a measure of how fast the system is cooled. The feedback parameter $\eta$ has the unit $\text{Time/Length}^2$, which can be tuned to control the speed of cooling and the final steady state energy.

\end{appendix}

\bibliographystyle{ieeetr}
\bibliography{all.bib}

\end{document}